\documentclass{aa}  

\usepackage{natbib}             
\bibpunct{(}{)}{;}{a}{}{,}

\usepackage{caption}            
\usepackage{graphicx}
\usepackage{multirow}
\usepackage{txfonts}
\usepackage{lipsum}
\usepackage{siunitx}
\usepackage{subcaption}         
\usepackage{lscape}             
\usepackage{placeins}           

\usepackage{rotating}
\usepackage{booktabs}
\usepackage{threeparttable}
\usepackage{adjustbox}
\usepackage[colorlinks=true, linkcolor=blue, citecolor=blue, urlcolor=blue]{hyperref}

\begin{document}

\title{The Alma catalog of OB stars}
\titlerunning{A massive star census of the Magellanic Clouds System}
\subtitle{IV. A massive star census of the Magellanic Clouds System}

%
%
%

\author{J. A. Molina-Calzada\inst{1,2}\email{jmolina@cab.inta-csic.es}
        \and J. Maíz Apellániz\inst{1}
        \and M. Pantaleoni González\inst{3}}

\institute{Centro de Astrobiología, CSIC-INTA, Campus ESAC, C. bajo del Castillo s/n, E-\num[detect-all]{28692}, Villanueva de la Cañada, Madrid, Spain
           \and Departamento de Física de la Tierra y Astrofísica, Universidad Complutense de Madrid, E-\num[detect-all]{28040} Madrid, Spain
           \and University of Vienna, Department of Astrophysics, Türkenschanzstrasse 17, 1180 Wien, Austria}
           
\date{Received 17 September 2026 / Accepted 23 November 2026}

 
\abstract
{The Magellanic Clouds System (MCS), comprising the Large and Small Magellanic Clouds (LMC and SMC) and the Magellanic Bridge (MB), provides an excellent laboratory for studying massive stars in the nearby extragalactic environment. However, existing catalogs are often limited by extinction, source crowding, and a lack of cross-validation with the literature.}
{We aim to identify massive star candidates in the MCS, defined here as stars with initial masses $M_{\rm i}\gtrsim8~M_\odot$, including O- and early-B-type stars, supergiants (SGs), luminous blue variables (LBVs), and Wolf--Rayet (WR) stars. The catalog is designed to improve completeness, particularly in highly reddened and crowded regions, and to recover candidates absent from large-scale surveys.}
{We combine \textit{Gaia} DR3 astrometry and photometry with SIMBAD, \textit{HST}, and complementary optical and infrared surveys. Candidate selection is based on color--magnitude and color--color diagrams, extinction tracks, \textit{Gaia} astrometry and variability diagnostics, spectroscopic classifications, and the treatment of source duplications and multiplicity. We additionally include literature sources and \textit{HST}-resolved objects in regions, such as the crowded 30~Dor, lacking reliable \textit{Gaia} counterparts. Synthetic \textit{Gaia} photometry is derived for sources with incomplete \textit{Gaia} photometry using complementary data.}
{The ALS~IV catalog contains \num{51954} massive star candidates, including \num{36506} in the LMC, \num{15264} in the SMC, and \num{184} in the MB. The catalog combines homogeneous identifiers and coordinates with multi-wavelength photometry, astrometry, variability information, and available spectral classifications. Complementary photometry and \textit{HST} observations allow us to recover reddened sources and stars in crowded regions missed by \textit{Gaia}-based selections. The sample traces the main populations of massive stars and prominent star-forming regions across the Clouds and the Bridge.}
{ALS~IV provides the most comprehensive census of massive star candidates in the MCS, addressing several limitations of previous catalogs. It provides a reference for studies of massive stars and a foundation for future surveys and \textit{Gaia} releases.}


\maketitle
\nolinenumbers

\section{Introduction} \label{sec:Introduction}

The "Alma\footnote{The name Alma refers to Alma College in Michigan, where the project’s originator was based, and has no relation to the Atacama Large Millimeter/submillimeter Array (ALMA) observatory.} Luminous Stars" (ALS) catalog \citep{Reed03} (ALS~I) originated more than three decades ago in the work of \citet{Reed93a} as a compilation of Galactic luminous stars with available \textit{UBV$\beta$} photometric data. Its foundation lay in the Case-Hamburg survey of Galactic luminous stars \citep{Hardetal59, Stocetal60, NassStep63, Hardetal64, Hardetal65, Nassetal65, StepSand71}, which provided the initial large-scale spectrophotometric material. In parallel, a similar initiative, the Galactic O-Star catalog (GOSC) \citep{Maizetal04b, Sotaetal08}, was launched by some of us two decades ago. Originally restricted to O-type stars, GOSC was later broadened to include massive stars in general\footnote{\url{https://gosc.cab.inta-csic.es}}. 

After a period of relative dormancy, the ALS project was revitalized with the advent of \textit{Gaia}. In \citet{Pantetal21} (ALS~II), \textit{Gaia} DR2 enabled the first cross-match with the original ALS~I data, providing additional information on the sources, including \textit{Gaia} photometry and astrometry. This allowed mapping massive stars in the solar neighborhood and led to the identification of a new Galactic structure, the Cepheus spur. With the release of \textit{Gaia}~DR3, \citet{Pantetal25b} (ALS~III) refined these results, expanding the sample, improving distance estimates, incorporating further information from the literature, and strengthening several of the catalog’s definitions.

In ALS~IV, we aim to extend our analysis beyond the Milky Way (MW) by incorporating the Magellanic Clouds System (MCS), which consists of the Large Magellanic Cloud (LMC), the Small Magellanic Cloud (SMC), and the Magellanic Bridge (MB). The MCS provides an excellent laboratory for massive star studies: its location above the Galactic plane mitigates extinction effects \citep{Gorsetal20}, while its extensive young population and interactions with the MW enhance massive star formation \citep{Rezaetal14}. 

Previous efforts to catalog massive stars in the MCS ranged from early works \citep[e.g.,][]{Sand70, Azzoetal75} that identified and classified high-luminosity stars as probable MCS members, to later studies \citep[e.g.,][]{Evanetal04, Bonaetal09} that provided spectral types and multiwavelength photometry for hundreds or thousands of objects. Large-scale surveys such as VISTA Magellanic Clouds \citep[VMC;][]{Cionetal11b}, \textit{Gaia} \citep{Prusetal16}, or One Thousand and One Magellanic Fields \citep[1001MC;][]{Cionetal19} further expanded the samples by millions, enabling new selection methods for the most recent catalogs \citep[e.g.,][]{Lurietal21, Schoetal26}. However, most studies focus on the upper part of the Color-Magnitude Diagram (CMD), leaving the more ambiguous regions underexplored \citep[e.g.,][]{MartHump23, Nakaetal25}. The alternative approach in this paper is to combine \textit{Gaia} astrometry and photometry with optical+NIR photometry and spectroscopy from other surveys to identify even moderately dust-extincted massive stars, carefully cross-matching the different samples to achieve a high degree of completeness, thus mitigating catalog issues from previous studies. In addition, we systematically incorporate objects reported only in the literature and absent from major survey catalogs, ensuring the most complete census possible.

In this work, we define massive stars as objects with initial masses $M_{\rm i}\gtrsim8~M_\odot$. This includes O stars; early-B-type main-sequence (MS) and giant stars; blue, yellow, and red supergiants (BSGs, YSGs, RSGs); luminous blue variables (LBVs); and Wolf--Rayet stars (WR). This definition is consistent with the classical definition of OB stars proposed by \citet{Morg51} and is based on the minimum initial mass required to produce a supernova \citep{Smar09}.

This paper is organized as follows. In Section~\ref{sec:Data and Methods}, we describe the datasets used to construct the catalog, and we present the methodology adopted to identify massive star candidates. In Section~\ref{sec:Results}, we describe the content of the final ALS~IV catalog. Finally, in Section~\ref{sec:Future Work}, we present the possible future research routes using and expanding the catalog.

\begin{figure*}
    \centering
    \includegraphics[width=\textwidth]{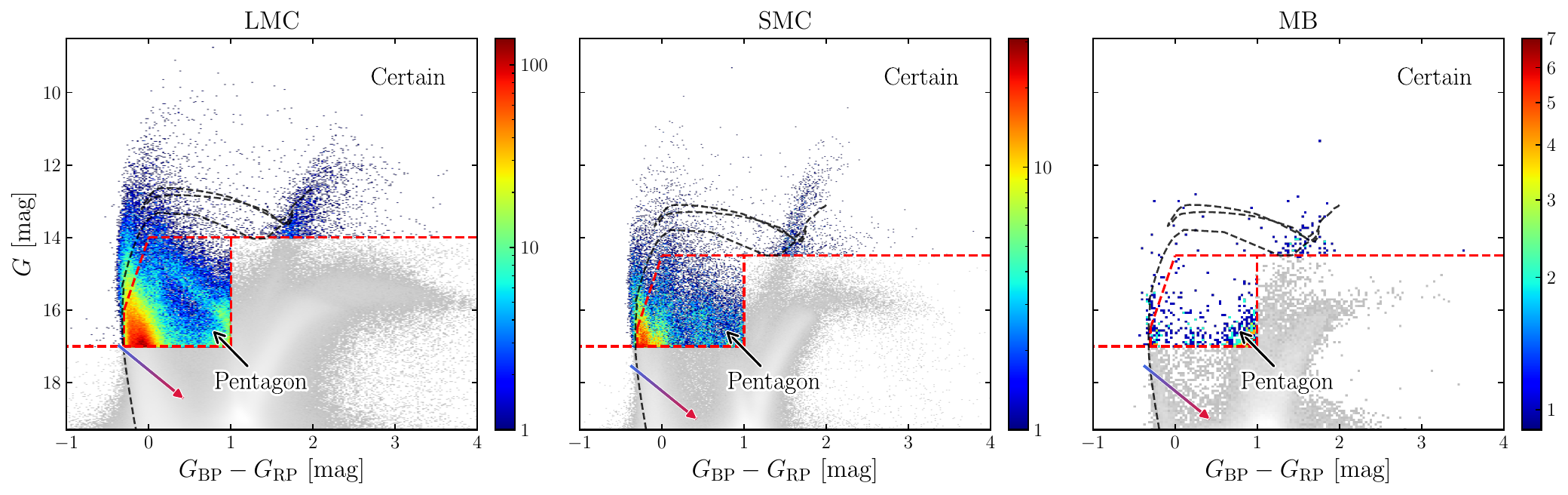}
    \caption{Photometric cut in the CMD for the LMC (left), SMC (center), and MB (right) for the \textit{Gaia}~DR3 sources filtered by \citet{Lurietal21}, including the selected sources from \citet{Schoetal26}. Colored and gray density points represent massive star candidates and discarded objects, respectively. The $G$-band magnitude thresholds are shown by the red dashed lines. The central polygon defines the Pentagon region, which requires further filtering as it lies below the $\sim8~M_\odot$ isochrone (without extinction; black dashed line), while the Certain region defines the initial candidate sample. The colored arrows show the effect of extinction (E(4405\,–\,5495) = 0.0\,–\,0.5) on an $\sim8~M_\odot$ ZAMS star \citep{Maizetal14a}.}
    \label{fig:photometric_cut_plot}
\end{figure*}

\section{Data and Methods} \label{sec:Data and Methods}

\subsection{Data sources} \label{sec:Data sources}

Different sources of information are used to create our catalog, with the two main ones being \textit{Gaia}~DR3 \citep{Valletal23} and SIMBAD \citep{Wengetal00}. The former provides astrometry and optical photometry, which are crucial for selecting MCS members and likely massive stars based on their positions in the CMD and astrometry. The latter is also fundamental, as it provides literature information, including not only additional photometry and astrometry but also spectral classifications, identifiers, and other relevant information. However, the information in the literature is manually reviewed before it is included in our catalog to minimize errors, which are frequent in SIMBAD.

In addition to these sources, we also include optical and NIR photometry. For the first one, we use the Magellanic Clouds Photometric Survey \citep[MCPS;][]{Zarietal02, Zarietal04}, which presents \textit{UBVI} photometry for the central regions of the LMC and SMC, and the SkyMapper Southern Sky Survey \citep[SkyMapper;][]{Onkeetal24}, covering an even wider region including the MB for \textit{uvgriz} bands. Regarding NIR photometry, we incorporate \textit{JHK} photometry from the Two Micron All Sky Survey \citep[2MASS;][]{Skruetal06} and \textit{YJKs} photometry from the VISTA survey of the Magellanic Clouds \citep[VMC;][]{Cionetal11b}, which provide complementary coverage, with 2MASS well suited to the brighter sources and VMC extending to fainter magnitudes.

Finally, crowding and multiplicity are observational challenges that are relevant in the MCS. The former is more significant in the 30~Dor region, where \textit{Gaia} photometry is incomplete and complementary surveys with higher resolution must be taken into account. Therefore, we use optical photometry from the Hubble Tarantula Treasury Project \citep[HTTP;][]{Sabbetal16}, obtained with the Hubble Space Telescope (HST) using the WFC3/UVIS and ACS/WFC instruments. We also incorporate spectroscopic classifications of stars in R136 from \citet{Crowetal16}, based on HST/STIS observations. Multiplicity (spectroscopic or visual) is a more serious issue: widely separated pairs may be resolved by \textit{Gaia}, but in most cases it may be identified only through information available in the literature, such as entries in SIMBAD that report a source as multiple.

\subsection{Philosophy} \label{sec:Philosophy}

The combination of multiple surveys represents a major advantage of this catalog over previous compilations, which were often restricted to a single data source. By combining datasets with different observational limitations, we reduce the impact of the biases inherent to any individual survey and improve both the completeness and reliability of the final sample. At the same time, the availability of multi-wavelength photometry is also crucial for identifying reddened massive stars, which are usually omitted when selecting from catalogs.

On the other hand, we incorporate information available in the literature to ensure that no relevant objects are overlooked. This is particularly important for sources with poor-quality photometry or astrometry, or those absent from surveys such as \textit{Gaia}, but previously identified and studied by other authors. In parallel, literature data also help identify multiplicity issues, as systems appearing as a single source in some surveys may have been resolved into multiple components in higher-resolution observations. The process is time-consuming but necessary for completeness.

\subsection{Initial data sets} \label{sec:Initial data sets}

The first step is to download the data from \textit{Gaia}~DR3 and SIMBAD for the MCS, composed of the LMC, SMC, and the MB. We define the LMC as a circular area of 10 degrees centered at $(81.28^\circ, -69.78^\circ)$ \citep{vadMCion01}, excluding the region defined as the MB below. Similarly, the SMC is defined as a circular area of 10 degrees centered at $(12.80^\circ, -73.15^\circ)$ \citep{Cionetal00b}, also excluding the region defined as the MB. We define the MB as a circular area of 10 degrees centered at $(43.5^\circ, -73.5^\circ)$ \citep{Skowetal14}, from which the innermost 7 degrees of the LMC and the innermost 6 degrees of the SMC are excluded. Note that this definition of the Bridge is more restrictive than those adopted in previous studies, which often include the SMC tidal arm \citep[e.g.,][]{Schmetal20, Schoetal26}.

The download criteria were as follows. For SIMBAD, we retrieved all sources with available coordinates (independently of magnitude or color) and then applied a filter to keep only stellar objects, based on the \texttt{otype}. This resulted in \num{322560} sources for the LMC, \num{191007} for the SMC, and \num{31067} for the MB in the SIMBAD samples, as of May 2025. For the \textit{Gaia}~DR3 data, we required that the downloaded sources have available coordinates, a parallax value, and $G<20.5$. Afterwards, a filtering procedure was applied following the method described in Section~2.2 of \citet{Lurietal21}, which is based on astrometric and photometric cuts. This filtering selected only sources with a high probability of belonging to the MCS according to astrometric and photometric criteria, removing foreground and background objects, as well as poor-quality sources. In addition, for completeness, we also included the \textit{Gaia} sources identified as MB members by \citet{Schoetal26} that were rejected by the selection method of \citet{Lurietal21} and not found in SIMBAD. These sources were retained based on the robustness of the supervised machine-learning classification performed by \citet{Schoetal26}; see Section~3 of that work for further details. Finally, we applied a photometric cut to the resulting CMD. After all these steps, we obtained \num{102911} sources for the LMC, \num{30377} for the SMC, and \num{657} for the MB for the \textit{Gaia} samples.

Figure~\ref{fig:photometric_cut_plot} shows the photometric cut applied to the CMD for the \textit{Gaia}~DR3 samples, filtered using the criteria of \citet{Lurietal21} in addition to \citet{Schoetal26} sources. This selection is based on $G$-band magnitude thresholds, with $G<17$ for $(G_\mathrm{BP}-G_\mathrm{RP})\leq1.0$ and $G<14$ and $G<14.5$ for $(G_\mathrm{BP}-G_\mathrm{RP})>1.0$ in the LMC and SMC/MB, respectively. These thresholds (red dashed lines) were chosen to include all massive stars (those above the $\sim8~M_\odot$ isochrone; black dashed line) without extinction and the majority of those with the typical low-to-moderate extinctions of the MCs. The colored arrow shows the effect of extinction on a $\sim8~M_\odot$ ZAMS star, from $E(4405-5495)=0.0$ to $0.5$, adopting the extinction law from \citet{Maizetal14a}. Distance modulus (DM) was also accounted for (DM = 18.5, 18.9, and 18.7\,mag for the LMC, SMC, and MB, respectively; \citealp{Pietetal19, Gracetal20, Jacyetal16}), but especially for redder stars, for which the magnitude limit is 0.5\,mag brighter in the LMC than in the SMC/MB. For bluer stars, we adopted a common limit of $G<17$ to ensure completeness in variability information from \citet{Maizetal23} and to maintain reliable \textit{Gaia} photometry. While these limits may exclude some highly extinguished massive stars, particularly in the more distant SMC and MB, some are recovered through SIMBAD (see Section~\ref{sec:simbad_only}), with further improvements expected from the future extension to \textit{Gaia}~DR4 (see Section~\ref{sec:Future Work}). As a result, the CMD is divided into different zones: an upper zone, which we call ``Certain'', expected to contain certain massive stars (from young and hot massive stars on the left to evolved and cool massive stars toward the right), and a central zone we call ``Pentagon'', which includes likely massive and intermediate-mass stars (from reddened main-sequence stars on the left side to blue-loop stars in the diagonal upper part). Including optical surveys with a $U$-band filter will help distinguish these populations, a key difference from previous catalogs, and help ensure that extincted massive stars are not overlooked.

\subsection{Selection of likely massive stars} \label{sec:Selection of likely massive stars}

\begin{figure*}
    \centering
    \includegraphics[width=\textwidth]{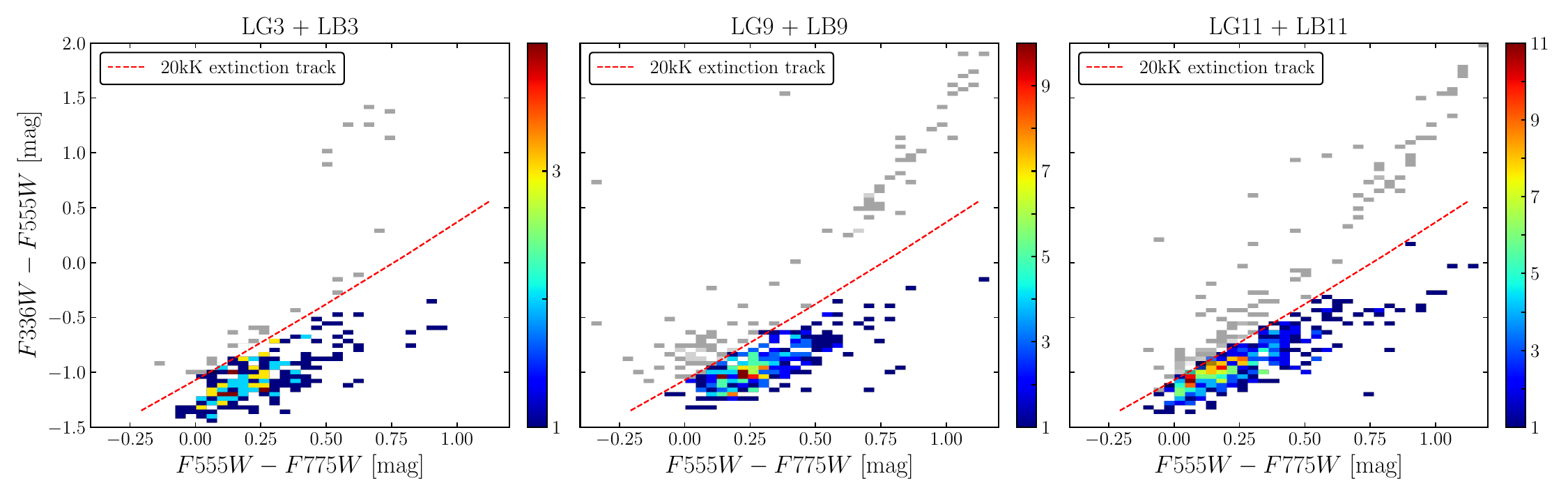}
    \caption{Color--color diagram of \textit{HST} photometry from the HTTP survey \citep{Sabbetal16} in the 30~Dor region. The left panel corresponds to the LG3 $+$ LB3 sample (\textit{Gaia} sources from the Pentagon region), whereas the center and right panels correspond to the LG9 $+$ LB9 and LG11 $+$ LB11 samples, which contain sources from SIMBAD-only and \textit{HST}-only, respectively. Colored points represent the selected sources (LG3, LG9, and LG11), while gray points indicate the excluded sources (LB3, LB9, and LB11). The selection is based on the extinction track derived from \citet{Maizetal14a} and \citet{MaizBarb18}.}
    \label{fig:hst_selection}
\end{figure*}

Once the initial \textit{Gaia}~DR3 and SIMBAD samples for the LMC, SMC, and MB are prepared, it is time to cross-match them. This procedure yielded three subsamples: sources present in both \textit{Gaia} and SIMBAD, sources found only in \textit{Gaia}, and sources found only in SIMBAD. The first two groups have available \textit{Gaia} photometry and astrometry (Sections \ref{sec:certain} and \ref{sec:pentagon}). For the latter, we initially only have information from SIMBAD (Section~\ref{sec:simbad_only}).

As a third source, in addition to \textit{Gaia} and SIMBAD, we incorporate HST photometry from the HTTP survey \citep{Sabbetal16} and spectroscopic information from \citet{Crowetal16} to mitigate the effects of crowding in the 30~Dor region of the LMC, as previously mentioned. This results in a fourth subsample for which only Hubble photometric data are available (Section~\ref{sec:hst_only}). The remaining surveys provide complementary information used to identify and characterize massive stars.

The entire methodology used to select massive stars is summarized in the flow charts shown in Figures~\ref{fig:flow_chart_lmc}, \ref{fig:flow_chart_smc}, and \ref{fig:flow_chart_mb} for the LMC, SMC, and MB, respectively. We recommend reading them alongside the corresponding sections in order to follow the details of the procedure more clearly.

\subsubsection{\textit{Gaia} sources in the Certain region} \label{sec:certain}

Starting with the sources for which \textit{Gaia} information is available (with or without a SIMBAD cross-match), an initial selection of unambiguously massive stars can be performed by selecting objects located in the upper part of the CMD, i.e., those within the Certain region, which includes both young main-sequence and evolved massive stars. This first selection constitutes the XG1 sample\footnote{Here, X = L for the LMC, X = S for the SMC, and X = B for the MB}, yielding an initial selection of \num{14456} massive stars for the LMC (LG1 sample), \num{7592} for the SMC (SG1 sample), and \num{145} for the MB (BG1 sample). Among the selected stars, we find HDE \num{269006} in the LMC, which is the brightest variable object in our sample, with $G \sim 8.7$ \citep[see][]{Walbetal17}.

The region named Certain is shown in Figure~\ref{fig:photometric_cut_plot}, with its boundaries defined to include all stars lying above the $\sim8~M_\odot$ isochrone (black dashed line). These isochrones are based on the three-dimensional grid of stellar models presented by \citet{Maiz13a}, which combines the Geneva \citep{LejeScha01} and Padova \citep{Giraetal00, Salaetal00} stellar evolutionary models. While the boundary is not perfect (e.g., some degeneracy between luminous AGBs and low-luminosity RSGs; see \citealp{Davietal18b}), it provides a conservative separation, minimizing the potential contamination. Therefore, based on their position in the CMD, we can identify these sources as massive stars with high confidence. This step represents the minimum selection criterion for identifying such objects and is the level of selection adopted in many studies \citep[e.g.,][]{Macketal17, Sunetal17, Nakaetal25}.

\subsubsection{\textit{Gaia} sources in the Pentagon region} \label{sec:pentagon}

The central polygon region of the CMD, which we define as the Pentagon (see Figure~\ref{fig:photometric_cut_plot}), is the most challenging, as it contains a mixture of extincted massive stars that have been reddened and therefore shifted toward the lower-right regions of the CMD, together with intermediate-mass stars such as some late-B and blue-loop stars. The right boundary of the Pentagon was chosen to encompass MS hot stars affected by up to $E(G_{\rm BP}-G_{\rm RP}) \sim 1.4$~mag of extinction and to avoid as much as possible the numerous intermediate-mass Asymptotic Giant Branch (AGB) population in the MCS. To remove these intermediate-mass contaminants and retain only the sources of interest, we make use of spectroscopic classifications when available and reliable, as well as variability information from \textit{Gaia} and optical photometry data from \textit{HST}, MCPS, and SkyMapper where applicable. The following steps are followed:

\begin{enumerate}
\setlength{\itemsep}{0.5em}

    \item Spectral type selection: The most reliable way to identify massive stars in this highly contaminated region is through their spectral classification. To this end, we make use of several spectroscopic surveys available in the literature. For the LMC, we consider spectral classifications from studies such as \citet{Breyetal99}, \citet{Evanetal15b}, or \citet{Crowetal16}. For the SMC and MB, we use classifications from works including \citet{Vinketal23} or \citet{Shenetal24a}. This selection allows us to recover \num{578} massive stars from the Pentagon region for the LMC (LG2 sample), \num{389} for the SMC (SG2 sample), and \num{1} for the MB (BG2 sample). Among the selected objects, we identify moderately extincted O-type stars, such as VFTS~117, with $G \sim 16.6$ \citep[see][]{Evanetal11a}.

    \item Variability exclusion: After the spectral-type selection, we further remove contaminating intermediate-mass variable stars using \textit{Gaia}~DR3 variability diagnostics \citep{Maizetal23}. In particular, we apply a two-branch criterion designed to identify classical pulsators (e.g., non-massive Cepheids and RR Lyrae), which typically exhibit both high-amplitude variability and a strong color dependence in their photometric variability. Sources are flagged as variable contaminants if they show either (i) large variability in the blue band ($S_{G_{\mathrm{BP}}} > 80$) combined with a blue-dominated variability signature ($S_{G_{\mathrm{BP}}} \gtrsim 1.25 \cdot S_{G_{\mathrm{RP}}}$), or (ii) a significant excess of blue over red variability ($S_{G_{\mathrm{BP}}} - S_{G_{\mathrm{RP}}} \gtrsim 0.6$) together with high $G$-band variability ($S_G > 60$). This procedure removes several thousand variable sources from the MCS (XB2 samples).

    \begin{figure*}
        \centering
        \includegraphics[width=0.85\textwidth]{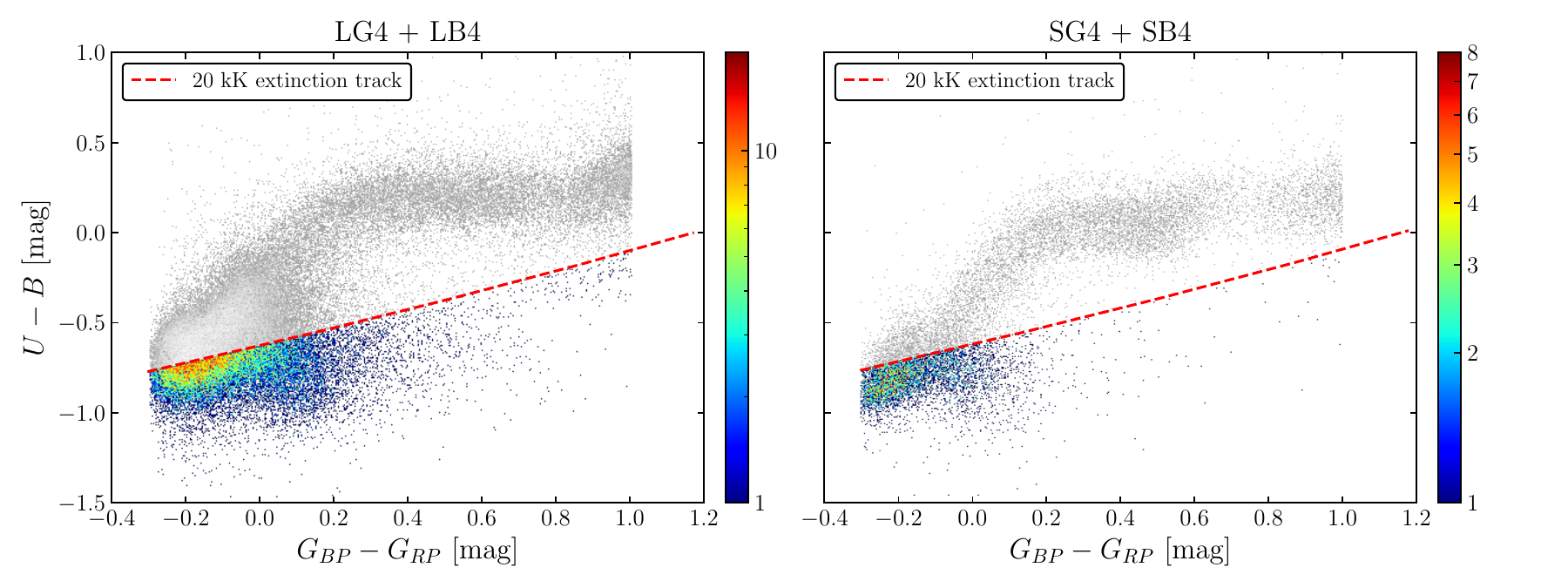}
        \caption{Color--color diagram of \textit{Gaia} sources cross-matched with MCPS \citep{Zarietal02, Zarietal04} for the LMC (left) and the SMC (right). Colored points represent the selected sources (XG4 samples), while gray points indicate the excluded sources (XB4 samples). The selection is based on  the extinction track derived from \citet{Maizetal14a} and \citet{MaizBarb18}.}
        \label{fig:zh_selection}
    \end{figure*}

    \item Cross-matches with optical surveys: After the two previous selection steps, we further refine the sample using additional optical photometric surveys aimed at identifying massive stars. Optical photometry that includes a $U$-band filter has long been used to select OB stars based on color-color diagrams such as $U-B$ vs. $B-V$ \citep{JohnMorg53}, but the procedure can be affected by calibration or extinction-related issues \citep{MaizSota08, Maizetal14a}. Here, our approach will be to use color-color diagrams with either (a) \textit{Gaia} $G_{\rm BP}-G_{\rm RP}$ (with its calibration of almost unparalleled quality, \citealt{MaizWeil25}) and a $U-B$-like filter from a ground-based survey on the $y$ axis or (b) \textit{HST} photometry \citep{Maizetal14a}.  The selection is defined using extinction tracks computed for a 20\,kK stellar model (corresponding to an initial mass of $\sim8~M_\odot$), adopting the extinction law of \citet{Maizetal14a} and $R_{5495} = 3.0$ from \citet{MaizBarb18}.

    For the LMC case, we first cross-match our catalog with \textit{HST} 30~Dor sources from the HTTP survey \citep{Sabbetal16}. This initial step is motivated by the severe crowding in this region, particularly around its central region R136, where many stars have visual companions at separations below 0.2\,arcsec \citep{Khoretal17}. Such close pairs are often unresolved in lower-resolution surveys, leading to source confusion and unreliable photometry. The superior angular resolution of space-based telescopes, such as \textit{HST}, is therefore essential for obtaining reliable source identifications in this region.

    Figure~\ref{fig:hst_selection} shows the \num{196} selected stars for 30~Dor (LG3 sample; left panel) obtained using the extinction track criteria in the color--color diagram, constructed using the \textit{HST} F336W (WFC3), F555W (ACS), and F775W (WFC3/UVIS, with ACS/WFC as fallback) bands. We find that nearly all sources are selected by our method, highlighting the high density of massive stars in the 30~Dor region. Some of the selected objects include the Op star identified as [ST92]~1-58 \citep[see][]{Walbetal14}.

    For the remaining objects, and to reintroduce the SMC into our analysis, we cross-match our sources with MCPS \citep{Zarietal02, Zarietal04}. This survey allows us to construct $U-B$-like versus \textit{Gaia} $G_{\rm BP}-G_{\rm RP}$ color--color diagrams. By projecting the extinction tracks onto this new parameter space, Figure~\ref{fig:zh_selection} shows the \num{19783} selected stars for the LMC (LG4 sample) and \num{5675} for the SMC (SG4 sample). Due to the limited sky coverage of this survey, the MB cannot be included in this step. Using this method, stars such as the Be star OGLE~J051918.40$-$694718.6 have been selected \citep[see][]{Saboetal05}.

    Lastly, to include the outer regions of the LMC and SMC together with the MB in our optical-band study, we incorporate photometric data from SkyMapper \citep{Onkeetal24}. In this case, we do not restrict ourselves to one color as done in previous steps. To maximize the number of usable sources given the survey characteristics, we additionally consider both the $u-v$ and $u-g$ colors. Specifically, we first apply the selection criteria in a color--color diagram using the extinction track based on $u-g$, and then, for sources lacking reliable $g$-band photometry, we apply the $u-v$ diagram when the photometric quality allows it. 
    
    Figure~\ref{fig:smss_selection} shows the corresponding color--color diagrams and extinction tracks, together with the selected samples using either $u-g$ or $u-v$ versus the \textit{Gaia} color. In this way, we select \num{293} stars for the LMC (LG5 sample), \num{1370} for the SMC (SG5 sample), and \num{31} for the MB (BG5 sample). This additional survey is particularly useful for probing the outer regions and significantly increases the sample size, especially for the SMC and MB. In this way, Bridge stars such as the gravity-mode pulsator [DI91]~1290 have been selected \citep[see][]{Aertetal23}.

    \begin{figure*}
        \centering
        \includegraphics[width=\textwidth]{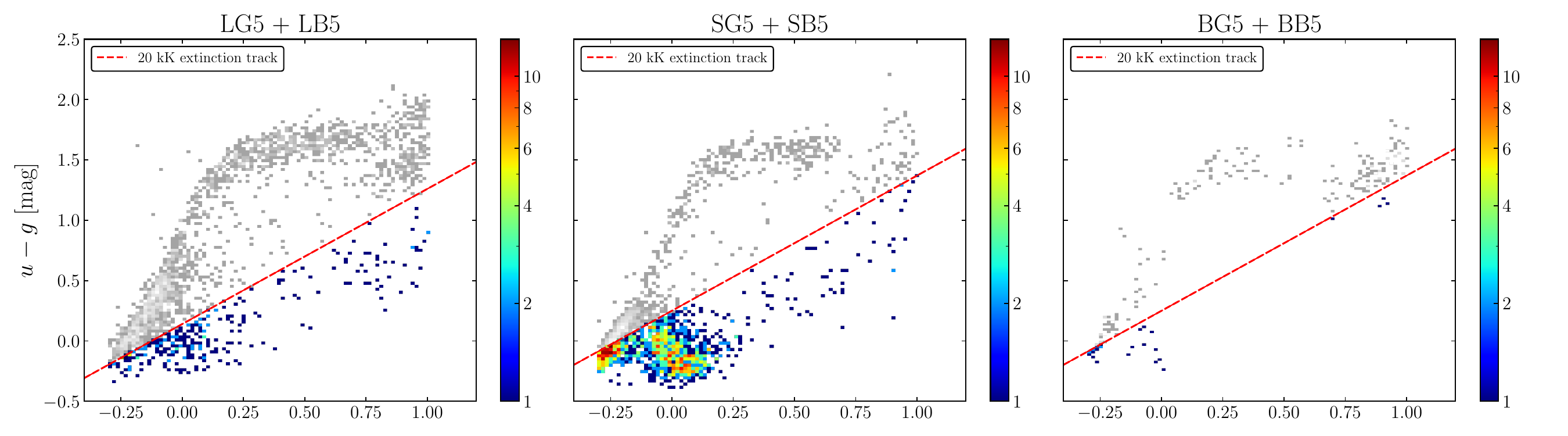}
        \includegraphics[width=\textwidth]{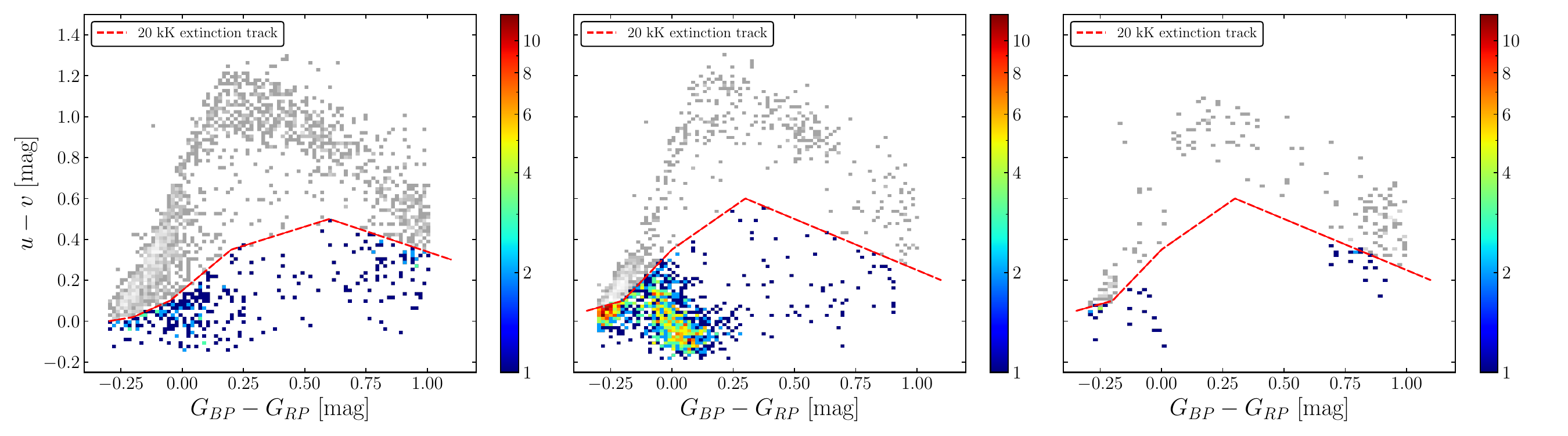}
        \caption{Color--color diagram of \textit{Gaia} sources cross-matched with SkyMapper \citep{Onkeetal24} for the LMC (left), SMC (center), and MB (right). Colored points represent the selected sources (XG5 samples), while gray points indicate the excluded sources (XB5 samples).  The selection is based on the extinction track derived from \citet{Maizetal14a} and \citet{MaizBarb18}. Depending on the availability and quality of the photometry, we apply either the $u-g$ (top) or the $u-v$ (bottom) color.}
        \label{fig:smss_selection}
    \end{figure*}

    All optical surveys used in this section were first filtered using survey-specific quality criteria. For the MCPS, we required valid $U$ and $B$ photometry ($U$ and $B$ $>$ 0), $\texttt{Flag}<20$, and removed color outliers. For SkyMapper, we retained only sources with clean photometric flags (\texttt{x\_flags} and \texttt{x\_nimaflags} $=$ 0), multiple good measurements (\texttt{x\_ngood} $>$ 1), no clipped or image-quality flags (\texttt{x\_nclip} $=$ 0), reliable PSF photometry (\texttt{flags\_psf} $=$ 0), stellar morphology ($\texttt{class\_star}>0.8$), and removed color outliers. For the \textit{HST} catalog, we required good-quality photometric flags ($f \leq 1$) in the relevant filters, excluded faint sources ($F555W \leq 18$), removed color outliers, and removed duplicate detections.

\end{enumerate}

\subsubsection{SIMBAD-only} \label{sec:simbad_only}

From the cross-match between \textit{Gaia} and SIMBAD, a subsample of sources is identified as present only in SIMBAD. Consequently, no \textit{Gaia} counterpart is initially available for these objects. Their identification therefore relies first on spectroscopic classifications, followed, in the LMC, by a photometric selection based on the \textit{HST} photometry from the HTTP survey \citep{Sabbetal16}. The remaining massive star candidates are then identified through a series of additional steps combining \textit{Gaia} astrometry and photometry, optical surveys, and information available in the literature. In practice, however, this procedure was applied separately to sources with identifiers from the main massive star catalogs of the MCS \citep[e.g.,][]{Feasetal60, Sand68, Azzoetal75, Brunetal75, Evanetal04, Evanetal15b}, bright sources ($B$, $V$, or $G < 17$), and faint sources or those lacking photometric measurements, to simplify the analysis. Nevertheless, for clarity, we present the complete procedure below as a single workflow:

\begin{enumerate}
\setlength{\itemsep}{0.5em}

    \item Spectral type: A first filtering step to identify massive star candidates is to select sources based on their spectral type. Following again the main spectroscopic studies mentioned, complemented by additional spectroscopic studies \citep[e.g.,][]{ReidPark12, Gonzetal15, Ramaetal19, Ishietal23, Ricketal24, Aranetal25}, \num{330} stars are selected for the LMC (LG7 sample) and \num{211} for the SMC (SG7 sample). For the MB, no sources can be selected under this criterion due to the lack of spectroscopic studies in this region. 
    
    An example of a star selected through this method is AzV~435, an O3V((f))z+? star \citep[see][]{Bestetal25} with a RUWE value of $\sim$12, which causes it to be excluded from many catalogs based on \textit{Gaia} astrometric quality criteria.

    \item Cross-match with \textit{HST}: Sources for the LMC are cross-matched with the HTTP catalog of 30~Dor from \citet{Sabbetal16}. This cross-match is performed because many objects located in these crowded regions cannot be resolved by \textit{Gaia} owing to its angular resolution limitations \citep{Khoretal17, Holletal23b}. In contrast, they are resolved in surveys of this kind (see Section \ref{sec:hst_only} for more detail). Figure~\ref{fig:hst_selection} presents the \num{344} selected stars (LG9 sample; middle panel), identified using the extinction track criteria applied to the color--color diagram constructed from the \textit{HST} F336W (WFC3), F555W (ACS), and F775W (WFC3/UVIS, with ACS/WFC as fallback) bands. Nearly all sources satisfy our selection criteria, reflecting again the exceptionally high concentration of massive stars in the 30~Dor region. 
    
    One example of an object selected by this method is the O7 star named CPD-69~457B \citep[see][]{Castetal21}, which has a RUWE of $\sim$6, so that were it not for the inclusion of \textit{HST} in the study, it would not have been included.

    \item Astrometry: We cross-matched the remaining sources with \textit{Gaia}. Notice that any counterparts recovered at this stage correspond either to sources excluded by \citet{Lurietal21} or to sources rejected by the photometric cut (see Figure~\ref{fig:photometric_cut_plot}). In the former case, the photometric and astrometric quality of the \textit{Gaia} measurements must be carefully assessed, since \citet{Lurietal21} excluded sources because of either poor astrometric solutions or likely foreground/background contamination. In the latter case, the recovered counterparts are expected to include contaminating populations, such as AGB stars, which must also be identified and removed.

    This cross-match provides parallaxes and proper motions for approximately 90\% of the sources in this subset. These measurements allow us to reject clear foreground and background contaminants by requiring parallaxes consistent with the expected values of each MCS (within 3--4$\sigma$, depending on the region, and with $\sigma_\varpi \leq 0.1$\,mas; \citealt{Lurietal21}) together with proper motions within conservative limits ($-2 \lesssim \mu_{\alpha},\mu_\delta \lesssim 5$\,mas\,a$^{-1}$; \citealt{Kalletal13, Schmetal20}), thereby rejecting approximately 48\% of the sample.

    For the remaining 10\% of the sources, for which no \textit{Gaia} astrometric information is available, it is not possible to reliably determine whether they belong to the MCS, unless their membership is supported by information from the literature.

    \item \textit{Gaia} photometry: For sources with available \textit{Gaia} astrometry and photometry and not excluded by astrometry, we examined their position in the CMD, selecting those located within the Certain region. Pentagon sources cannot be considered reliable massive star candidates due to the ambiguity. This procedure allowed us to recover \num{38} sources from the MCS. One of the selected objects is HD~\num{269215}A, which has $G\sim$ 12.3, and due to its likely multiplicity, it has a RUWE of $\sim$11.

    \item Variability: In this case, the variability criterion was identical to that described in Section~\ref{sec:pentagon}, using the variability classifications from \citet{Maizetal23} and applying the same selection criteria. This process removes approximately 7\% of the sample, many of these objects being Cepheids and RR Lyrae stars.

    \item Optical photometry: For the remaining sources with \textit{Gaia} astrometry and photometry values, we searched for counterparts in MCPS \citep{Zarietal02, Zarietal04} and SkyMapper \citep{Onkeetal24}, obtaining matches for approximately 30\% and 12\% of the sample, respectively. The corresponding photometric quality flags were taken into account as explained previously. Since the foreground and background contaminants have already been removed through the astrometric selection, the location of these sources in the color--color diagrams (see Figures~\ref{fig:zh_selection} and \ref{fig:smss_selection}) can be used to identify more massive star candidates. This procedure allowed us to recover \num{34} sources. One of the selected objects is OGLE~SMC-ECL-7665, an eclipsing binary with $U-B = -0.9$ \citep[see][]{Zarietal02}.

    \item Duplications and multiplicity: One of the most time-consuming aspects of compiling the catalog was identifying duplicated entries and accounting for source multiplicity. First, the same source is often cataloged under different identifiers in different catalogs without cross-identification, leading to multiple entries in the literature despite having angular separations of $\lesssim 0.1''$ and nearly identical astrometry and photometry. For example, Cl*~NGC 2070~SMB~30 and Brey~79 are listed as different objects in SIMBAD at the time of writing, although their positions and photometry clearly indicate that they correspond to the same source, as shown in \citet{Doraetal13}. Similarly, 2dFS~992 and ISO-MCMS~J005424.4$-$730335 are treated as different objects in the literature despite having identical coordinates. These are only two examples among many encountered during the creation of the catalog. Although SIMBAD progressively resolves such inconsistencies, they are still present in some cases and therefore require careful manual inspection to produce a catalog that is as complete and free of duplications as possible. Duplicated objects are recorded in the XD8 samples, whereas the outcome of the merger is reflected in the XG8 or XB8 samples, depending on whether the duplicate was accepted or rejected, respectively.
    
    A second challenge is source multiplicity. Depending on the angular resolution and source definition adopted by each catalog, what is considered a single object in one catalog may be resolved into multiple components in another. For example, RMC~140 is resolved into the components VFTS~507 (RMC~140a) and VFTS~509 (RMC~140b), whereas many catalogs identify the system as a single object (e.g., Brey~87 or HDE~\num{269919}). To further complicate matters, additional sources are found only a few arcseconds away, such as VFTS~519 (RMC~140c) and VFTS~497 (RMC~140d). Another example is VFTS~240, which actually corresponds to a system comprising \textit{Gaia}~DR3 4657680518305197824 and \textit{Gaia}~DR3 4657680518283581440. In this latter case, \textit{Gaia}~DR3 4657680518283581440 is itself classified by \citet{Gracetal11} as an eclipsing binary, illustrating that multiplicity issues may extend to hierarchical systems. This highlights a more general limitation: because of the finite angular resolution of existing observations and the incompleteness of current surveys, the true multiplicity of many massive stars remains unresolved \citep{Walbetal99b, Sanaetal12a}. Consequently, this issue cannot yet be fully addressed when constructing a homogeneous catalog. 
    
    This study has helped us include about \num{5} more massive star candidates (2 of them lacking \textit{Gaia} counterparts) and exclude many others. One example is [ST92]~2-62B, component B of the binary system [ST92]~2-62 \citep[see][]{SchiTest92}. Its component A was selected directly from the Certain part of the CMD, but its component B, having poor-quality photometry, was not studied through that route.

    \item Other factors: We also considered additional information available in the literature, such as the object classification and the catalog identifier. Objects already classified as AGB stars \citep[e.g.,][]{Bicaetal08} or white dwarfs \citep[e.g.,][]{Gentetal21} can be safely excluded based on their spectral classification. Likewise, some catalog identifiers correspond to sources projected onto the MCS but are physically unrelated to them. For example, objects associated with 47~Tucanae \citep[e.g.,][]{Tuch92} or NGC~1313 \citep[e.g.,][]{HadfCrow07} can be readily identified and excluded from the catalog. 
    
    Following this approach, and considering objects included in previous studies of the MCS \citep[e.g.,][]{MaluHeap94, Heydetal99, Massetal00, Crowetal16, Bodeetal20b}, \num{43} objects are selected as massive star candidates, 34 of which lack \textit{Gaia} counterparts. There are several examples drawn from the cluster R136, such as the star [HSH95]~86 \citep{Crowetal16} or BI~148 from the cluster LHA 120-N~127A \citep{Brunetal75}.

\end{enumerate}

Through this procedure, we recover \num{86} massive star candidates in the LMC (LG8 sample), \num{27} in the SMC (SG8 sample), and \num{7} in the MB (BG8 sample). Although this selection procedure may raise concerns about the reliability of the selected sources, the careful extraction process described in the preceding steps, systematically excluding any uncertain objects, ensures the purity of the sample. On the other hand, approximately 5\% of the sample remains unclassified by this process, as these objects lack sufficient information to draw reliable conclusions about their nature. Therefore, they are included as part of the XB8 exclusion samples. These objects include, for example, sources without \textit{Gaia} parallaxes or literature references, as well as objects from catalogs affected by spurious detections, either due to inaccurate coordinates or data-processing issues. 

\subsubsection{HST-only} \label{sec:hst_only}

As previously discussed, \textit{Gaia} suffers from limited angular resolution \citep{Holletal23b}, which is particularly problematic in crowded regions such as 30~Dor \citep{MoffSegg83, Khoretal17}, where multiple sources may remain unresolved. We therefore complement the existing sample with massive star candidates identified from \textit{HST}-resolved sources in dedicated studies of 30~Dor \citep{Crowetal16, Sabbetal16}. Other massive star-forming complexes, such as NGC~346 and N11, were also assessed using HST data \citep[e.g.,][]{Miloetal23}. However, crowding in these regions is not severe enough to compromise the \textit{Gaia} analysis \citep{Lurietal21}. Including these regions would add only a handful of stars to the final sample, with negligible impact on the results, while unnecessarily complicating the flow diagrams. Thus, this \textit{HST}-based selection is restricted to 30~Dor.

Two main \textit{HST} studies for 30~Dor are taken into account. The first study is \citet{Crowetal16}, which focuses on the innermost $\sim$0.85\,pc of 30~Dor, corresponding to the R136 cluster, where the high stellar density makes high-resolution HST/STIS spectroscopy particularly well suited to characterizing individual stars. This study also incorporates archival HST/WFC3 photometry from \citet{DeMaetal11a}. This region contains the vast majority of the detected massive stars, including R136a1, the most massive star known in the cluster \citep{Crowetal10}. The second study is the HTTP survey from \citet{Sabbetal16}, which covers a significantly larger field, encompassing the full 30~Dor complex. This catalog covers a field of $\sim$210 $\times$ 180\,pc in eight \textit{HST} filters, with a relative astrometric accuracy of 0.4\,mas, making it particularly well suited to identifying individual sources across the crowded 30~Dor region.

By combining the sources from \citet{Crowetal16} and the HTTP survey, applying the same quality filtering criteria based on photometric flags as those used for the \textit{HST} data in Section~\ref{sec:pentagon}, and removing sources already present in our initial \textit{Gaia}+SIMBAD sample, we obtain \num{645} new 30~Dor objects. To avoid duplicate entries, we additionally verified that no object from \citet{Crowetal16} was already included in the HTTP survey, and vice versa. Among these \num{645} sources, \num{8} objects are part of the sample of \citet{Crowetal16}. Given their location within the R136 core and their confirmed massive star nature, these sources are directly included in our catalog as the LG10 sample.

The remaining \num{637} sources from the HTTP survey are subjected to the same photometric selection procedure described in Section~\ref{sec:pentagon}. Specifically, their photometry is examined in the color--color diagram shown in the right panel of Figure~\ref{fig:hst_selection}, where we use the extinction tracks from \citet{Maizetal14a} and \citet{MaizBarb18} to identify sources consistent with massive stars. Although we adopt a uniform $R_{5495} = 3.0$, despite the variations in the extinction law reported in 30~Dor by \citet{Maizetal14a}, we do not expect this assumption to significantly affect our selection, since the impact is greatest for highly extinguished sources, which represent only a small fraction of the population. Therefore, this analysis yields \num{432} additional massive star candidates, which constitute the LG11 sample. This result further highlights the exceptional richness of 30~Dor in massive stars and the importance of accounting for highly crowded regions.

Combining all these sources, we finally obtain a sample of \num{980} massive star candidates (LG3 + LG9 + LG10 + LG11 samples) in the 30~Dor region, complementing previous censuses of massive stars \citep[e.g.,][]{Doraetal13, Walbetal14}.

\subsection{Catalog description} \label{sec:catalog description}

The result of the cleaning and classification process is presented in the form of a catalog listing the \num{51954} massive star candidates selected. Table~\ref{tab:catalog} describes the information included in each column. An online version will be available on our website and at \texttt{VizieR} for download (see \hyperref[sec:data_availability]{Data Availability} section).

The ALS~IV catalog provides a comprehensive census of massive stars in the MCS together with homogeneous identifiers (\texttt{id\_als}) that facilitate its use and future extensions. Following the nomenclature philosophy introduced in ALS~III, we adopt region-specific prefixes: LLS (LMC Luminous Star), SLS (SMC Luminous Star), and BLS (Bridge Luminous Star) for sources in the LMC, SMC, and MB, respectively. For the LMC and SMC, identifiers are assigned within predefined $G$ synthetic magnitude ranges and ordered by increasing $BP-RP$ synthetic color within each range. The gaps between consecutive identifier blocks provide room for future additions without requiring the renumbering of existing sources. MB sources are numbered sequentially in increasing $G$ synthetic magnitude. Sources subsequently identified as false positives may result in gaps in the sequences in the future. The adopted numbering scheme is summarized in Table~\ref{tab:als_nomenclature}. Overall, approximately 98\% of the catalog sources are estimated to be genuine massive star candidates.

\begin{table}[ht]
\renewcommand{\arraystretch}{1.15}
\centering
\caption{Identifier numbering scheme adopted for the ALS~IV catalog.}
\label{tab:als_nomenclature}
    \begin{tabular}{lll}
        \hline \hline
        Region & $G$ range & Identifier start \\
        \hline
        \multirow{7}{*}{LMC}
            & 8--12  & LLS \num{1} \\
            & 12--13 & LLS \num{1001} \\
            & 13--14 & LLS \num{3001} \\
            & 14--15 & LLS \num{8001} \\
            & 15--16 & LLS \num{14001} \\
            & 16--17 & LLS \num{25001} \\
            & $>17$  & LLS \num{40001} \\
        \hline
        \multirow{6}{*}{SMC}
            & 9--13  & SLS \num{1} \\
            & 13--14 & SLS \num{1001} \\
            & 14--15 & SLS \num{3001} \\
            & 15--16 & SLS \num{6001} \\
            & 16--17 & SLS \num{10001} \\
            & $>17$  & SLS \num{19001} \\
        \hline
        MB & -- & BLS \num{1} \\
        \hline
    \end{tabular}
\end{table}

As a consequence, 7 sources inherited from the original ALS catalog were found to belong to the MCS rather than to the MW population. These sources are ALS~15\,895 (LLS~155), ALS~15\,896 (LLS~104), ALS~18\,185 (LLS~1\,200), ALS~18\,840 (LLS~1\,017), ALS~18\,845 (LLS~1\,013), ALS~19\,598 (LLS~205), and ALS~19\,597 (SLS~178). To avoid duplicate identifiers, we adopt the corresponding LLS/SLS identifiers in the present catalog, while retaining the original ALS identifiers as cross-identifications.

We also provide the equatorial coordinates of each source as right ascension (\texttt{ra}) and declination (\texttt{dec}) in the ICRS/J2000 reference frame. When multiple positional measurements are available, we adopt a hierarchical priority of HST, \textit{Gaia}, and SIMBAD. HST positions are preferred in crowded regions owing to their higher angular resolution, followed by \textit{Gaia} astrometry and, when neither is available, SIMBAD coordinates.

Regarding the \textit{Gaia} photometry, we provide the corrected $G$ magnitude (\texttt{g\_mag\_corr\_gaia}) and photometric uncertainties (\texttt{g\_mag\_error\_gaia}, \texttt{bp\_mag\_error\_gaia}, and \texttt{rp\_mag\_error\_gaia}), following \citet{MaizWeil25}, which accounts for systematic effects in the \textit{Gaia} photometric calibration. We also provide corrected parallaxes (\texttt{parallax\_corr\_gaia}, \texttt{parallax\_error\_corr\_gaia}) and proper motions (\texttt{pmra\_corr\_gaia}, \texttt{pmra\_error\_corr\_gaia}, \texttt{pmdec\_corr\_gaia}, and \texttt{pmdec\_error\_corr\_gaia}), accounting for known systematic biases in \textit{Gaia} astrometry, following \citet{CanGBran21}, \citet{Lindetal21b}, and \citet{Maiz22}. These corrections were applied whenever the necessary \textit{Gaia} information was available. Further details on the correction procedures and recipes are available in the associated GitHub repository\footnote{\url{https://github.com/MichelangeloPantaleoni/Gaia-DR3-Corrections-and-recipes}}.

As a complement to the \textit{Gaia} photometry, the catalog provides synthetic \textit{Gaia} photometry for all sources. Although approximately 99\% of the sources have complete \textit{Gaia} photometry, about 1\% are missing one or more of the three \textit{Gaia} magnitudes, mainly because they are located in highly crowded regions (e.g., 30~Dor) or were included based on their spectral type. For these sources, complementary photometry from other surveys was used to derive synthetic \textit{Gaia} photometry. The resulting $G$ magnitude and $G_{\rm BP}-G_{\rm RP}$ color are provided in \texttt{g\_syn} and \texttt{bp\_rp\_syn}, respectively, while \texttt{g\_source} and \texttt{bp\_rp\_source} indicate the photometric data used in their derivation. When \textit{Gaia} photometry was unavailable, complementary data were selected following the priority order \textit{HST} (WFC3, with WFPC2 as fallback), Johnson photometry (e.g., MCPS), SkyMapper, and 2MASS.

\section{Results} \label{sec:Results}

In this section we present our first results from the analysis of the MCS massive star population.

\subsection{Sample size}

The final ALS~IV sample contains \num{51954} massive star candidates: \num{36506} in the LMC, \num{15264} in the SMC, and \num{184} in the MB (Table~\ref{tab:sample_size}).

\begin{table}[ht]
    \renewcommand{\arraystretch}{1.15}
    \caption{Final number of massive star candidates identified in the ALS~IV catalog for the MCS.}
    \label{tab:sample_size}
    \centering
    \begin{tabular}{ll}
        \hline \hline
        Region & Candidates \\
        \hline
        LMC & \num{36506} \\
        SMC & \num{15264} \\
        MB  & \num{184} \\
        \hline
        Total & \num{51954} \\
        \hline
    \end{tabular}
\end{table}

\begin{figure*}
    \centering
    \includegraphics[width=\textwidth]{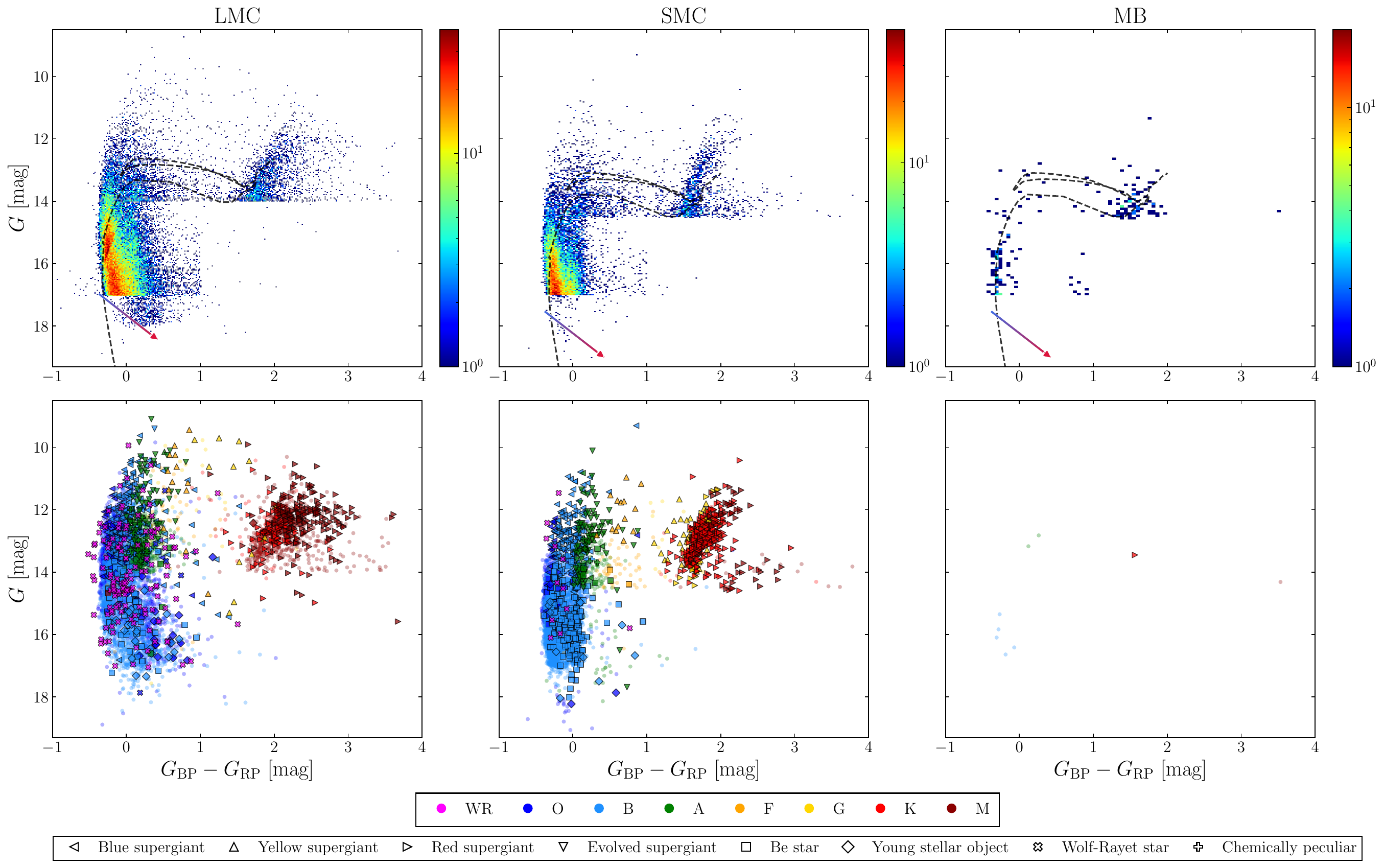}
    \caption{(Top) Synthetic CMDs of the LMC (left), SMC (center), and MB (right) for the final ALS~IV catalog of \num{51954} massive star candidates. Black dashed lines represent the $\sim8~M_\odot$ isochrone. (Bottom) The same CMDs for the subset of sources with SIMBAD spectral classifications. Colors indicate spectral type, while symbols represent the SIMBAD object type indicated when available. The colored arrows show the effect of extinction (E(4405\,–\,5495) = 0.0\,–\,0.5) on an $\sim8~M_\odot$ ZAMS star \citep{Maizetal14a}.}
    \label{fig:CMD_final}
\end{figure*}

\begin{figure*}
    \centering
    \includegraphics[width=\textwidth]{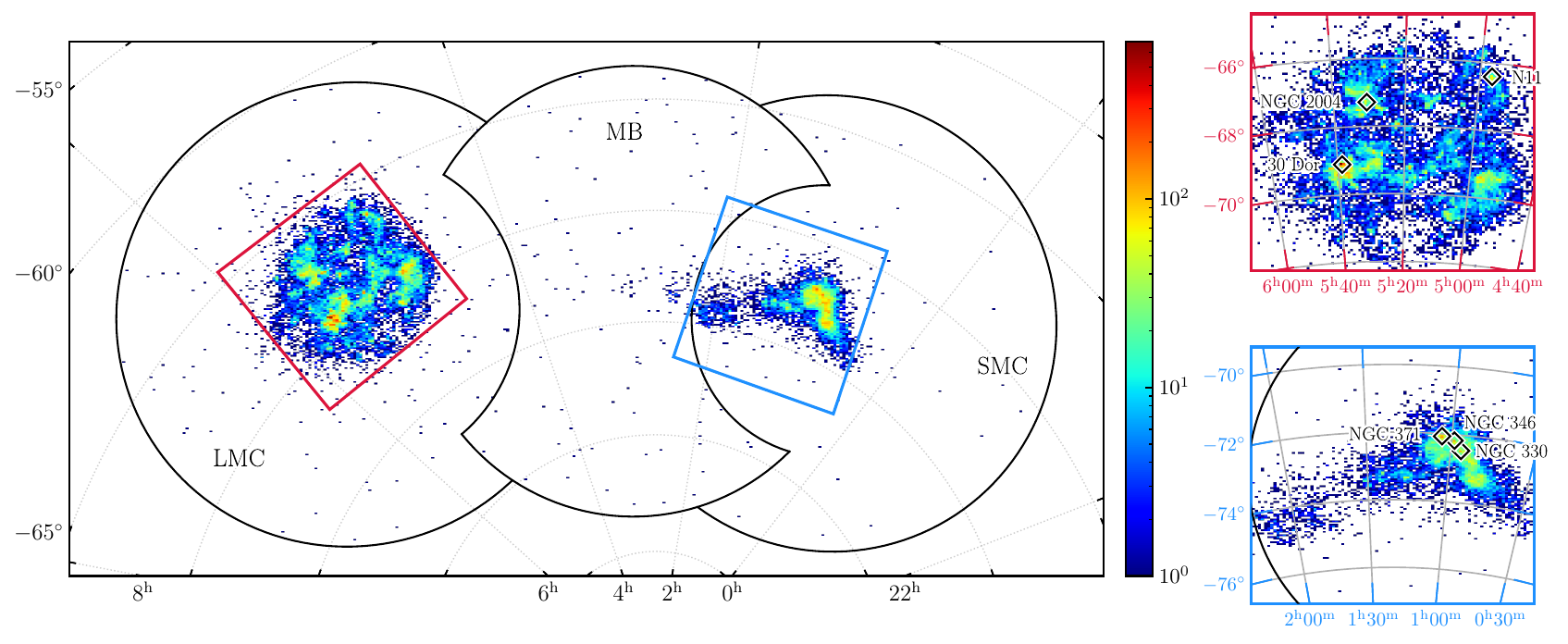}
    \caption{Spatial distribution of the final \num{51954} ALS~IV massive star candidates in the MCS. The map includes the LMC, SMC, and MB defined regions, showing the main structures traced by the massive star population. Red and blue quadrants zoom in on the central regions of the LMC and SMC, respectively. Some of the main OB associations are marked with diamonds.}
    \label{fig:sky_plot_limits}
\end{figure*}

The size of our sample differs from those reported in previous studies for two main reasons. First, differences may arise from the definition of a massive star itself. In our case, as discussed in the introduction, we define a massive star as an object with an initial mass above $\sim8~M_\odot$ \citep[see][]{Maizetal26a}. However, other studies may adopt different criteria. For instance, some broadly select the upper part of the CMD without imposing a specific lower-mass limit \citep[e.g.,][]{Poggetal21}. In contrast, others adopt a less restrictive mass threshold, considering objects with masses above $\sim2~M_\odot$ as massive \citep[e.g.,][]{Xianetal22}. Finally, some studies restrict their samples to massive stars with spectroscopic confirmation, resulting in substantially smaller sample sizes \citep[e.g.,][]{Xuetal21}.

Second, the size and boundaries of the analyzed regions must also be considered, particularly for the Bridge. Some studies define the Bridge region broadly, including a significant fraction or even the entire tidal halo of the SMC \citep[e.g.,][]{Skowetal14, Schoetal26}, whereas our definition is almost the opposite, with these outer regions being included within the SMC sample \citep[e.g.,][]{Castetal18b, Nakaetal25}. Similarly, some works focus exclusively on the central regions of the LMC and SMC, excluding their outer environments \citep[e.g.,][]{Zarietal02, Zarietal04}, while others cover several degrees of angular extent \citep[e.g.,][]{Lurietal21}.

This explains, for instance, the differences between our sample sizes and those reported in previous studies: \citet{Schoetal26} identified \num{12864} stars in the MB, compared to the \num{184} massive stars selected in this work, mainly due to the different definition of the Bridge region adopted in each study and the definition of a massive star. Similarly, \citet{Nakaetal25} reported \num{7426} massive stars in the SMC, whereas our catalog contains \num{15264} objects, as their selection does not include extincted massive stars. Finally, \citet{MartHump23} identified 3862 massive stars in the LMC using a more restrictive definition of massive star based only on spectral type, compared to the \num{36506} stars included in our sample.

\subsection{Color-Magnitude Diagrams}

\begin{figure*}
    \centering
    \includegraphics[width=\textwidth]{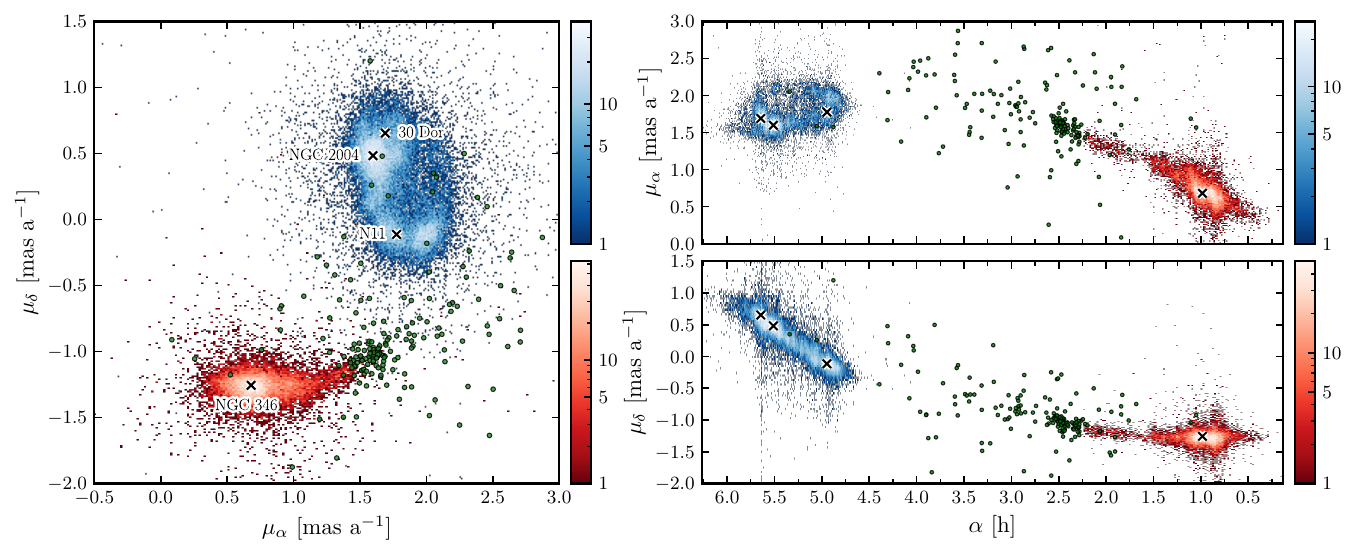}
    \caption{(Left) Distribution of the \textit{Gaia} proper motions for the final ALS~IV sample. The blue and red density distributions correspond to the LMC and SMC, respectively, while the green points correspond to the MB. Crosses indicate the positions of some main regions of the Clouds in the proper-motion space. Distinct internal structures are apparent within both the LMC and SMC, as well as the Bridge connecting the two Clouds. (Right) \textit{Gaia} proper motions as a function of right ascension. The connection between the two Clouds through the Bridge is again clearly apparent.}
    \label{fig:proper_motion_mc_sample}
\end{figure*}

Figure~\ref{fig:CMD_final} shows the CMDs of the final ALS~IV massive star candidate sample across the LMC, SMC, and MB. The upper panels show all \num{51954} sources in the catalog, using \textit{Gaia} magnitudes obtained via the synthetic photometry procedure described in Section~\ref{sec:catalog description}. The lower panels show only the subset of sources with spectral classifications available in SIMBAD, with the corresponding SIMBAD object types indicated when available. The upper CMDs recover the main stellar populations and evolutionary sequences expected for massive stars, including the prominent main sequence traced by OB stars and the parallel sequence associated with Be stars. The diagrams also reveal the very blue colors of WR stars and the regions of the CMD occupied by evolved massive stars, such as BSGs, RSGs, and YSGs. Moreover, the inclusion of the Pentagon region makes the direction of extinction-induced reddening clearly visible, with this effect being more pronounced in the LMC than in the SMC \citep[see][]{JoshPanc19}.

The effects of our selection process can be seen by comparing Figures~\ref{fig:photometric_cut_plot}~and~\ref{fig:CMD_final}. The decrease in the number of sources in the Pentagon region of the CMD is readily apparent, but the transition across the boundary with the Certain region is relatively smooth, as expected from an accurate selection process for massive stars. For the LMC and SMC, the stars that remain in the Pentagon are concentrated toward the left boundary, as expected from a massive MS population with small and varying degrees of extinction. Only a few extincted stars are seen with $G_{\rm BP}-G_{\rm RP}\,>\,0.5$~mag, and the structure that marks the blue end of the blue loop has disappeared altogether, as expected given it is made of intermediate-mass stars inside the Pentagon. The stars that are seen in Figure~\ref{fig:CMD_final} fainter than $G=17$ are those added as SIMBAD-only and \textit{HST}-only, with the current cut at that magnitude being a consequence of the limitations of \textit{Gaia}~DR3.

The lower CMDs in Figure~\ref{fig:CMD_final} include only sources with a spectral classification in SIMBAD, shown by color, and their SIMBAD object type, shown by symbol, when available. The LMC and SMC are both dominated by O- and B-type stars, although their relative contributions differ significantly: the LMC contains approximately 1700 O- and 2095 B-type stars, whereas the SMC contains 490 O- and 3257 B-type stars \cite[see][]{Garmetal87, Hilletal94b, Evanetal06}. In terms of the SIMBAD object types, the SMC contains substantially larger populations of cool supergiants, particularly RSGs, with around 940 objects compared to 274 in the LMC, and YSGs, with 119 compared to 28 \citep[see][]{MassOlse03, Neugetal10, OGraetal24}. Conversely, the LMC hosts a much larger population of WR stars, with about 152 objects classified as such compared to only 12 in the SMC \citep[see][]{MassDuff01, Hainetal15, Massetal15c}, as well as more Be stars, with 776 versus 406 in the SMC \citep[see][]{Saboetal05, Bonaetal10}. The LMC also contains approximately 65 chemically peculiar (including metal-poor) stars, in comparison with the 19 in the SMC \citep[see][]{Paunetal06, Vinketal23}. The MB sample is considerably smaller and has very sparse spectroscopic coverage, highlighting the need for further spectroscopic studies in this region \citep[see][]{Ramaetal21, Schoetal25}.

These numbers are indicative rather than complete, as they are based on SIMBAD information and depend on observational biases and heterogeneous spectral classifications. For instance, previous dedicated censuses of RSGs have identified larger populations in the LMC than in the SMC \citep[e.g.,][]{Renetal21}, in contrast to the numbers based on SIMBAD object types reported here. This discrepancy may be partly attributed to heterogeneous SIMBAD classifications. Similar completeness limitations also affect the O-star population. The spectroscopically confirmed O-type stars therefore represent only a subset of the O-star population included in ALS IV, since many additional candidates lack spectral classifications. Estimating the total number of O-type stars through further spectroscopic observations or population synthesis models \citep[e.g.,][]{Hawcetal25} accounting for the star formation history \citep[e.g.,][]{HarrZari04, HarrZari09} and stellar evolution of each galaxy, is left for future work. The different sizes of the LMC and SMC should also be considered when comparing these values.

\subsection{Spatial distribution and proper motions}

Figure~\ref{fig:sky_plot_limits} shows the spatial distribution of the final \num{51954} massive star candidates selected in the ALS~IV sample. Several well-known young stellar populations across the MCS are clearly identified, including 30~Dor and N11 in the LMC, or NGC~346 and NGC~330 in the SMC, as well as the extended tidal halo of the SMC connecting with the MB. The characteristic barred structure of the LMC is not clearly apparent in the distribution of massive stars, which instead exhibits a globally heterogeneous morphology with several enhanced star-forming regions along its pseudo-bars \citep{RuiLetal20}. In contrast, the SMC shows a more homogeneous spatial distribution, strongly affected by its interaction with the LMC and extending toward it through the Bridge \citep{Besletal16}.

Although each population has its own characteristic dynamics and kinematics, an overall gradient in proper motions across each Cloud ultimately defines its general internal structure \citep[see][]{Lurietal21}. This structure is apparent in Figure~\ref{fig:proper_motion_mc_sample}, which shows the proper-motion distribution in right ascension and declination for the MCS. For both the LMC and SMC, the proper motions are clearly concentrated in distinct regions of parameter space. For the LMC, the lower right panel of Figure~\ref{fig:proper_motion_mc_sample} reveals the characteristic correlation of a rotating system. In the left panel we can identify at least three internal overdensities associated with the spiral-like features and the bar, with the positions of several prominent massive star regions marked by crosses in proper-motion space. In contrast, the SMC exhibits a proper-motion gradient extending from a central overdensity (associated with NGC 346 and its surroundings) toward the region of parameter space occupied by the LMC, which reveals the stretching caused by the influence of the LMC, with the proper motions overlapping with the distribution of those of the MB. Thus, a clear connection is apparent not only in the spatial distribution of the stars, as shown in Figure~\ref{fig:sky_plot_limits}, but also in proper-motion space, where the two Clouds are connected through the Bridge. Therefore, these motions suggest that the SMC is being tidally stretched by the LMC and/or affected by ram pressure \citep[see][]{Zivietal19, Murretal24, Nakaetal25}.

Finally, a number of massive stars exhibit proper motions that deviate from the dominant kinematic patterns of the Clouds, while others are found in isolation in the outer regions. These objects may include runaway or isolated stars, which have been identified in the MCS as a result of dynamical interactions and/or binary evolution \citep[see][]{Gvaretal10, Gvaretal11a, Gvaretal12a}.

\section{Future Work} \label{sec:Future Work}

The most immediate improvement of our knowledge of the massive star sample of the MCS  will come from \textit{Gaia}~DR4. The increased astrometric precision and improved photometric calibrations will significantly enhance the identification of massive stars. In particular, the methodology developed here for the Pentagon region of the CMD can be extended approximately one magnitude deeper, from the current completeness limit of $G\sim17$ to $G\sim18$. Such an extension will allow the inclusion of a substantially larger population of early-B and more heavily reddened massive stars, reducing contamination near the selection boundaries while increasing completeness in the most extincted regions. We also considered including stripped stars \citep[e.g.,][]{Ludwetal26}, many of which lie below the current $G=17$ limit. However, since our current methods cannot robustly distinguish stripped stars that were originally massive from those that originated from intermediate-mass stars, we only include those confirmed as massive in the literature. Improved photometric information from \textit{Gaia}~DR4 may eventually enable a more complete census of these objects, particularly when combined with ultraviolet (UV) data \citep[e.g.,][]{Hotaetal24, Ludwetal26}. Furthermore, the introduction of Gaia's new crowded-field processing pipeline in DR4 is expected to substantially improve source detection and characterization in the densest regions \citep{Weinetal23}, where source blending currently limits the catalog completeness. This will be particularly relevant for regions such as 30~Dor. Looking further ahead, \textit{Gaia}~DR5 is expected to provide another substantial improvement thanks to the availability of the full mission data.

The ALS~IV catalog will also provide a unique opportunity to study the relationship between interstellar dust and massive star populations. Massive stars have been proposed as powerful tracers of the three-dimensional distribution and properties of the interstellar medium, particularly when combined with Galactic dust maps, revealing connections between young stellar populations and regions of enhanced extinction \citep{Lalletal22, Edenetal24}. The ongoing exploitation of ALS~III is exploring this approach in the MW, while ALS~IV will extend it to the MCS, allowing us to investigate whether similar trends are present in lower-metallicity environments.

The scientific legacy of ALS~IV will enable a wide range of studies of massive stars, including the spatial distribution of OB associations, the three-dimensional structure of the LMC and SMC, the stellar populations of the MB, or the impact of metallicity on massive star evolution. The catalog will also facilitate systematic investigations of runaway stars, evolved massive stars, highly reddened massive stars, RSGs, Be stars, variability, and multiplicity, among other topics. Besides, the Bridge provides a unique opportunity to identify extremely metal-poor massive stars under early-Universe-like conditions. Another important application will be to estimate the total O-type star population in the MCS using further spectroscopic observations and population synthesis models accounting for the star formation history \citep[e.g.,][]{HarrZari04, HarrZari09} and stellar evolution of each galaxy. This will help assess the completeness of the current census and better characterize the massive star populations of the LMC and SMC. These results will also provide an opportunity to better constrain the high-mass end of the initial mass function (IMF) and investigate possible differences between the two Clouds.

ALS~IV is expected to become an increasingly valuable community resource for both the MW and the MCS. Future versions of the catalog will provide carefully selected target lists for future spectroscopic observations, particularly in poorly studied regions such as the MB, while also serving as a training dataset for automated classification algorithms based on machine learning. In this way, ALS~IV is expected to evolve from a catalog into a reference database for the study of massive stars.

\section*{Data Availability}
\phantomsection
\label{sec:data_availability}

The full ALS~IV catalog, of which Table~\ref{tab:catalog_example_combined} shows an excerpt, is available electronically at the CDS via anonymous FTP at \url{cdsarc.u-strasbg.fr} (130.79.128.5) or via \url{http://cdsweb.u-strasbg.fr/cgi-bin/qcat?J/A+A/}. An interactive version of the catalog is also available at \url{https://juananmol.github.io/als-iv-stellar-atlas/}.

\begin{acknowledgements} \label{sec: Acknowledgments}

    We would like to thank B. Cameron Reed for initiating the ALS catalog project and for his pioneering work, which provided the foundation for the subsequent ALS catalogs. We acknowledge support from the Spanish Government Ministerio de Ciencia, Innovaci\'on y Universidades and Agencia Estatal de Investigaci\'on (AEI, \num{10.13039}/\num{501100011033}) through grant PID2022-\num{136640}~NB-C22. In addition, MPG was co-funded by the European Union (ERC, ISM-FLOW, 101055318) and acknowledges funding from the European Research Council (ERC) under the European Union’s Horizon 2020 research and innovation programme (Grant agreement No. 851435). We have made use of data from the European Space Agency (ESA) mission \textit{Gaia} (\url{https://www.cosmos.esa.int/gaia}), processed by the \textit{Gaia} Data Processing and Analysis Consortium (DPAC, \url{https://www.cosmos.esa.int/web/gaia/dpac/consortium}). Funding for the DPAC has been provided by national institutions, particularly those participating in the \textit{Gaia} Multilateral Agreement. This research has made use of the SIMBAD database, Aladin Sky Atlas, and the VizieR catalog access tool, operated and developed by the Centre de Données astronomiques (CDS), Strasbourg Observatory, France, as well as TOPCAT, an interactive graphical viewer and editor for tabular data. We mainly used the Python packages \texttt{astropy} \citep{Robietal13}, \texttt{astroquery} \citep{Ginsetal19}, \texttt{matplotlib} \citep{Hunt07}, and \texttt{numpy} \citep{Harretal20}.

\end{acknowledgements}

\bibliographystyle{bibtex/aa}
\bibliography{general}

\onecolumn
\begin{appendix}

    \section{Flow charts} \label{sec:Flow charts}

\begin{figure*}[h]
        \centering
        \includegraphics[width=0.83\textwidth]{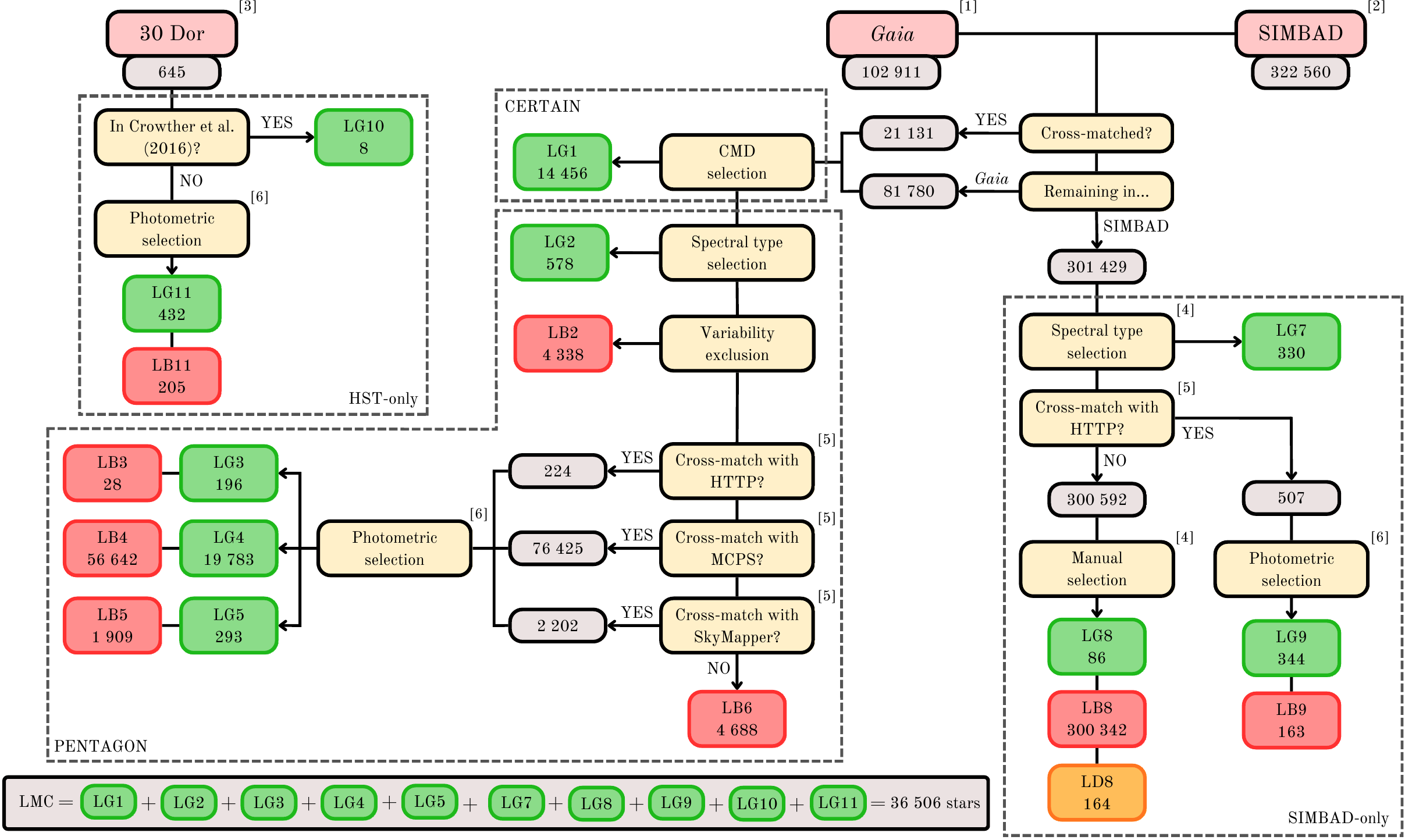}
        \caption{Flow chart of the LMC, showing the number of objects at each step. Sample labels are denoted as LX$N$, where X = G (selected; in green), B (rejected; in red), and D (duplicated; in orange), and $N$ is the step. [1] \textit{Gaia}~DR3 sample based on the filtering of \citet{Lurietal21} and \citet{Schoetal26}, with the photometric cut applied (Figure~\ref{fig:photometric_cut_plot}). [2] SIMBAD sample filtered by object type. [3] \citet{Crowetal16} and HTTP survey sources filtered by flags and not in [1] or [2] samples. [4] Selection based on literature references. [5] Survey filtered by flags. [6] Selection using the 20\,kK extinction track \citep{Maizetal14a}.}
        \label{fig:flow_chart_lmc}
    \end{figure*}

\begin{figure*}[h]
        \centering
        \includegraphics[width=0.83\textwidth]{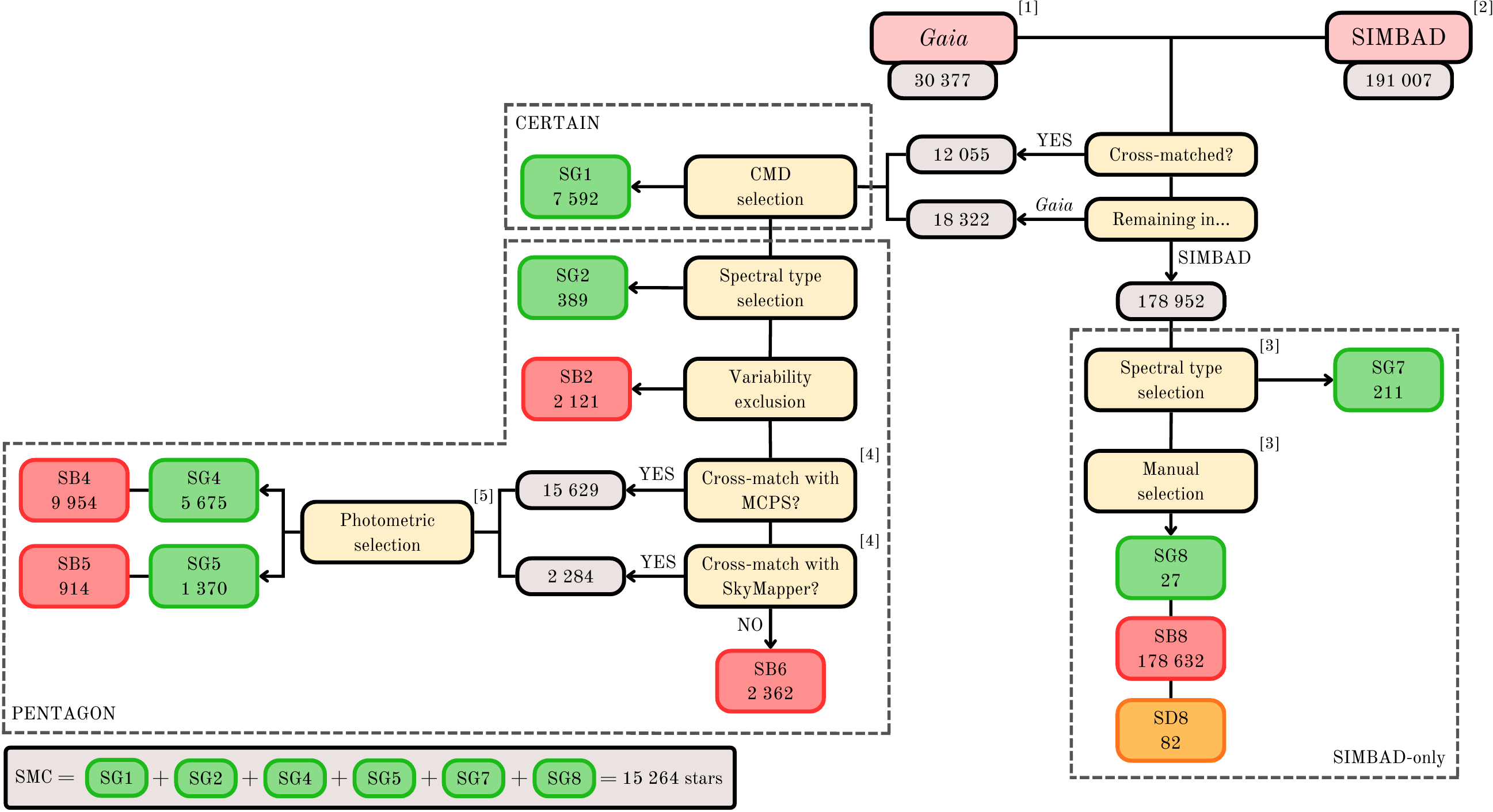}
        \caption{Flow chart of the SMC, showing the number of objects at each step. Sample labels are denoted as SX$N$, where X = G (selected; in green), B (rejected; in red), and D (duplicated), and $N$ is the step. [1] \textit{Gaia}~DR3 sample based on the filtering of \citet{Lurietal21} and \citet{Schoetal26}, with the photometric cut applied (Figure~\ref{fig:photometric_cut_plot}). [2] SIMBAD sample filtered by object type. [3] Selection based on literature references. [4] Survey filtered by flags. [5] Selection using the 20\,kK extinction track \citep{Maizetal14a}.}
        \label{fig:flow_chart_smc}
    \end{figure*}

\begin{figure*}[h]
        \centering
        \includegraphics[width=0.83\textwidth]{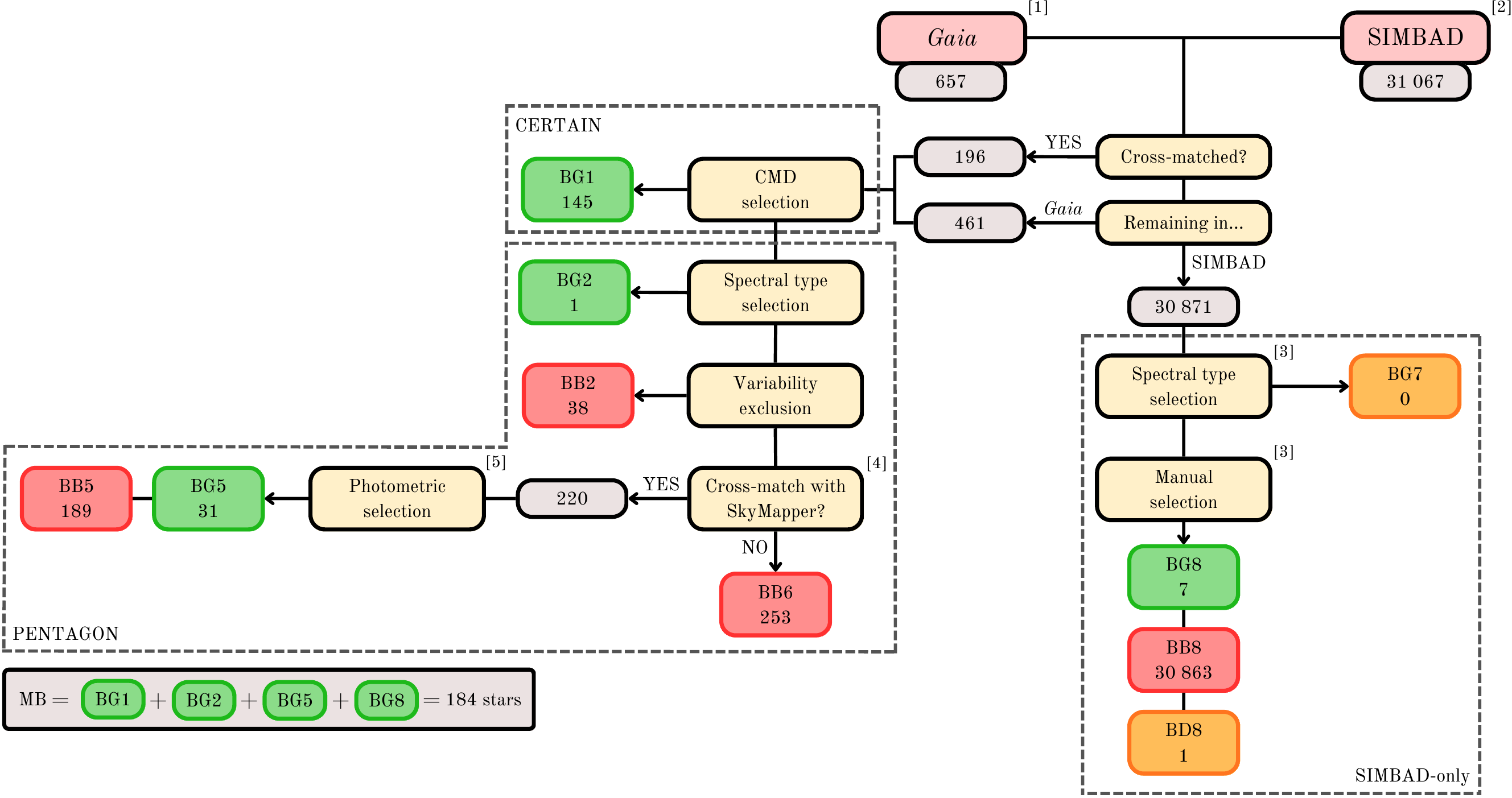}
        \caption{Flow chart of the MB, showing the number of objects at each step. Sample labels are denoted as BX$N$, where X = G (selected; in green), B (rejected; in red), and D (duplicated; in orange), and $N$ is the step. [1] \textit{Gaia}~DR3 sample based on the filtering of \citet{Lurietal21} and \citet{Schoetal26}, with the photometric cut applied (Figure~\ref{fig:photometric_cut_plot}). [2] SIMBAD sample filtered by object type. [3] Selection based on literature references. [4] Survey filtered by flags. [5] Selection using the 20\,kK extinction track \citep{Maizetal14a}.}
        \label{fig:flow_chart_mb}
    \end{figure*}

    Figures~\ref{fig:flow_chart_lmc}, \ref{fig:flow_chart_smc}, and \ref{fig:flow_chart_mb} show the flow charts summarizing the methodology adopted throughout this work to identify (in green) and reject (in orange and red) massive star candidates for the LMC, SMC, and MB, respectively.
    
    \section{Catalog} \label{sec:catalog}

    Table~\ref{tab:catalog} lists the columns included in the ALS~IV catalog of massive star candidates selected from the MCS, together with their units and description.
    
    \renewcommand{\arraystretch}{1.15}
    \begin{longtable}{lll}
    
    \caption{Description of the columns in the ALS~IV catalog.}
    \label{tab:catalog}\\
    \hline\hline
    Column name & Units & Description \\
    \hline
    \endfirsthead
    
    \caption{continued.}\\
    \hline\hline
    Column name & Units & Description \\
    \hline
    \endhead
    
    \hline
    \endfoot
    
    
    \multicolumn{3}{c}{\textbf{Identifiers \& Classification}}\\
    \hline
    
    id\_als & --- & ALS~Identifier\\
    id\_dr3\_gaia & --- & \textit{Gaia}~DR3 identifier\\
    main\_id\_simbad & --- & SIMBAD main identifier\\
    id\_simbad & --- & SIMBAD additional identifiers\\
    id\_hubble & --- & \textit{HST} identifier\\
    id\_2mass & --- & 2MASS identifier\\
    id\_vista & --- & VISTA identifier\\
    id\_smss & --- & SkyMapper identifier\\
    otype\_simbad & --- & SIMBAD object type\\
    sp\_type\_simbad & --- & SIMBAD spectral type\\
    sp\_qual\_simbad & --- & SIMBAD spectral type quality flag\\
    sp\_bibcode\_simbad & --- & SIMBAD bibcode for spectral type\\
    Sample & --- & Sample classification\\
    Notes & --- & Additional comments\\
    
    \hline
    
    
    \multicolumn{3}{c}{\textbf{Coordinates}}\\
    \hline
    
    ra & deg & Right ascension (ICRS/J2000; \textit{HST}$\rightarrow$\textit{Gaia}$\rightarrow$SIMBAD priority)\\
    dec & deg & Declination (ICRS/J2000; \textit{HST}$\rightarrow$\textit{Gaia}$\rightarrow$SIMBAD priority)\\
    
    \hline
    
    
    \multicolumn{3}{c}{\textbf{\textit{Gaia}~DR3 photometry} \citep{Valletal23}}\\
    \hline
    
    g\_mag\_gaia & mag & \textit{Gaia} $G$ magnitude\\
    g\_mag\_corr\_gaia & mag & Corrected \textit{Gaia} $G$ magnitude\\
    g\_mag\_error\_gaia & mag & Uncertainty in \textit{Gaia} $G$ magnitude\\
    bp\_mag\_gaia & mag & \textit{Gaia} $G_{BP}$ magnitude\\
    bp\_mag\_error\_gaia & mag & Uncertainty in $G_{BP}$ magnitude\\
    rp\_mag\_gaia & mag & \textit{Gaia} $G_{RP}$ magnitude\\
    rp\_mag\_error\_gaia & mag & Uncertainty in $G_{RP}$ magnitude\\

    g\_syn & mag & Synthetic \textit{Gaia} $G$ magnitude\\
    bp\_rp\_syn & mag & Synthetic \textit{Gaia} $G_{BP}-G_{RP}$ color\\
    g\_source & --- & Photometric source for synthetic $G$\\
    bp\_rp\_source & --- & Photometric source for synthetic $G_{BP}-G_{RP}$\\
    
    \hline
    
    
    \multicolumn{3}{c}{\textbf{\textit{Gaia}~DR3 astrometry} \citep{Valletal23}}\\
    \hline
    
    parallax\_gaia & mas & Parallax\\
    parallax\_corr\_gaia & mas & Corrected parallax\\
    parallax\_error\_gaia & mas & Parallax uncertainty\\
    parallax\_error\_corr\_gaia & mas & Corrected parallax uncertainty\\
    
    pmra\_gaia & mas/a & Proper motion in RA\\
    pmra\_corr\_gaia & mas/a & Corrected proper motion in RA\\
    pmra\_error\_gaia & mas/a & Proper motion uncertainty in RA\\
    pmra\_error\_corr\_gaia & mas/a & Corrected proper motion uncertainty in RA\\
    
    pmdec\_gaia & mas/a & Proper motion in DEC\\
    pmdec\_corr\_gaia & mas/a & Corrected proper motion in DEC\\
    pmdec\_error\_gaia & mas/a & Proper motion uncertainty in DEC\\
    pmdec\_error\_corr\_gaia & mas/a & Corrected proper motion uncertainty in DEC\\
    
    ruwe\_gaia & --- & Renormalized Unit Weight Error (RUWE)\\
    astrometric\_params\_solved\_gaia & --- & Astrometric solution flag\\
    
    \hline
    
    
    \multicolumn{3}{c}{\textbf{Variability from \textit{Gaia}~DR3} \citep{Maizetal23}}\\
    \hline
    
    SG\_var & mmag & Variability for $G$ band\\
    SSG\_var & mmag & Uncertainty in variability for $G$ band\\
    SBP\_var & mmag & Variability for $G_{BP}$ band\\
    SSBP\_var & mmag & Uncertainty in variability for $G_{BP}$ band\\
    SRP\_var & mmag & Variability for $G_{RP}$ band\\
    SSRP\_var & mmag & Uncertainty in variability for $G_{RP}$ band\\
    
    \hline

    
    \multicolumn{3}{c}{\textbf{HST photometry} \citep{Crowetal16, Sabbetal16}}\\
    \hline
    
    F336W\_hubble & mag & \textit{HST}/WFC3 F336W magnitude\\
    F555W\_hubble & mag & \textit{HST}/ACS F555W magnitude\\
    F775W\_hubble & mag & \textit{HST} F775W magnitude (WFC3/UVIS$\rightarrow$ACS/WFC priority)\\
    e\_F336W\_hubble & mag & Uncertainty in F336W\\
    e\_F555W\_hubble & mag & Uncertainty in F555W\\
    e\_F775W\_hubble & mag & Uncertainty in F775W\\
    f\_F336W\_hubble & --- & Photometry quality flag for F336W\\
    f\_F555W\_hubble & --- & Photometry quality flag for F555W\\
    f\_F775W\_hubble & --- & Photometry quality flag for F775W\\
    
    \hline
    
    
    \multicolumn{3}{c}{\textbf{Infrared photometry} \citep{Skruetal06, Cionetal11b}}\\
    \hline
    
    j\_2mass & mag & 2MASS $J$ magnitude\\
    h\_2mass & mag & 2MASS $H$ magnitude\\
    k\_2mass & mag & 2MASS $K_s$ magnitude\\
    j\_err\_2mass & mag & Uncertainty in $J$ band\\
    h\_err\_2mass & mag & Uncertainty in $H$ band\\
    k\_err\_2mass & mag & Uncertainty in $K_s$ band\\
    Qflag\_2mass & --- & 2MASS photometric quality flag\\
    
    Kspmag\_vista & mag & VISTA $K_s$ magnitude\\
    Jpmag\_vista & mag & VISTA $J$ magnitude\\
    Ypmag\_vista & mag & VISTA $Y$ magnitude\\
    e\_Kspmag\_vista & mag & Uncertainty in $K_s$\\
    e\_Jpmag\_vista & mag & Uncertainty in $J$\\
    e\_Ypmag\_vista & mag & Uncertainty in $Y$\\
    Kserrb\_vista & --- & Photometry quality flag for $K_s$\\
    Jerrb\_vista & --- & Photometry quality flag for $J$\\
    Yerrb\_vista & --- & Photometry quality flag for $Y$\\
    
    \hline
    
    
    \multicolumn{3}{c}{\textbf{MCPS photometry} \citep{Zarietal02, Zarietal04}}\\
    \hline
    
    U\_zh & mag & $U$ band magnitude\\
    B\_zh & mag & $B$ band magnitude\\
    V\_zh & mag & $V$ band magnitude\\
    I\_zh & mag & $I$ band magnitude\\
    
    e\_U\_zh & mag & Uncertainty in $U$\\
    e\_B\_zh & mag & Uncertainty in $B$\\
    e\_V\_zh & mag & Uncertainty in $V$\\
    e\_I\_zh & mag & Uncertainty in $I$\\
    
    Flag\_zh & --- & Photometric quality flag\\
    
    \hline
    
    
    \multicolumn{3}{c}{\textbf{SkyMapper photometry} \citep{Onkeetal24}}\\
    \hline
    
    u\_psf\_smss & mag & $u$ PSF magnitude\\
    v\_psf\_smss & mag & $v$ PSF magnitude\\
    g\_psf\_smss & mag & $g$ PSF magnitude\\
    r\_psf\_smss & mag & $r$ PSF magnitude\\
    i\_psf\_smss & mag & $i$ PSF magnitude\\
    z\_psf\_smss & mag & $z$ PSF magnitude\\
    
    e\_u\_psf\_smss & mag & Uncertainty in $u$\\
    e\_v\_psf\_smss & mag & Uncertainty in $v$\\
    e\_g\_psf\_smss & mag & Uncertainty in $g$\\
    e\_r\_psf\_smss & mag & Uncertainty in $r$\\
    e\_i\_psf\_smss & mag & Uncertainty in $i$\\
    e\_z\_psf\_smss & mag & Uncertainty in $z$\\
    
    u\_flags\_smss & --- & Photometric quality flag for $u$\\
    v\_flags\_smss & --- & Photometric quality flag for $v$\\
    g\_flags\_smss & --- & Photometric quality flag for $g$\\
    r\_flags\_smss & --- & Photometric quality flag for $r$\\
    i\_flags\_smss & --- & Photometric quality flag for $i$\\
    z\_flags\_smss & --- & Photometric quality flag for $z$\\
    
    \end{longtable}

\vspace{1em}
Table \ref{tab:catalog_example_combined} shows part of the catalog as an example. It includes only some columns and a selection of objects from the different ranges defined in Table \ref{tab:als_nomenclature} to give an idea of the different types of sources that can be found.

\begin{sidewaystable*}
\caption{Representative excerpt from the ALS~IV catalog. Three sources are listed for each LMC and SMC synthetic-magnitude interval, together with a representative selection of MB sources. The selection includes objects from both the principal and the less populated catalog samples.}
\label{tab:catalog_example_combined}
\centering
\scriptsize
\renewcommand{\arraystretch}{1.15}
\setlength{\tabcolsep}{3pt}
\begin{adjustbox}{max width=\textheight}
\begin{tabular}{llllllllrlrllrlrrrl}
\hline\hline
ALS ID & \textit{Gaia} DR3 ID & SIMBAD ID & R.A. & Dec. & Otype & SpT & $G_{\rm syn}$ &
$(G_{\rm BP}-G_{\rm RP})_{\rm syn}$ & Syn. source & $S_G$ & $U$ & UV source &
$K_s$ & IR source & $\varpi_{\rm corr}$ & $\mu_{\alpha,\rm corr}$ &
$\mu_{\delta,\rm corr}$ & Sample \\
 & & & [deg] & [deg] & & & [mag] & [mag] & $G$/color & [mmag] & [mag] & &
[mag] & & [mas] & [mas/a] & [mas/a] & \\
\hline
LLS~128 & 4662413606586588160 & HD~\num{270754} & 71.76855 & -67.11475 & s*b & B1.5Ia & 11.178 & 0.105 & \textit{Gaia}/\textit{Gaia} & $22.51 \pm 0.30$ & $10.351 \pm 0.104$ & ZH & $12.668 \pm 0.002$ & VMC & $-0.052 \pm 0.034$ & $0.684 \pm 0.046$ & $0.961 \pm 0.047$ & LG7 \\
LLS~198 & 4661472149794844288 & HD~\num{33579} & 76.48132 & -67.88637 & sg* & A2Ia+ & 9.097 & 0.338 & \textit{Gaia}/\textit{Gaia} & $27.29 \pm 0.06$ & $10.108 \pm 0.014$ & SM & $9.687 \pm 0.001$ & VMC & $0.025 \pm 0.027$ & $1.781 \pm 0.036$ & $0.140 \pm 0.039$ & LG1 \\
LLS~210 & 4658204053297963392 & HD~\num{269216} & 78.37828 & -69.53990 & Em* & OBe & 9.845 & 0.391 & \textit{Gaia}/\textit{Gaia} & $142.15 \pm 0.01$ & $10.227 \pm 0.099$ & ZH & $9.958 \pm 0.001$ & VMC & $-0.001 \pm 0.021$ & $2.027 \pm 0.034$ & $0.155 \pm 0.033$ & LG1 \\
LLS~1\,287 & 4658459900242918912 & SK~-68~122 & 83.33625 & -68.68983 & HXB & B0.5Ib & 12.691 & -0.099 & \textit{Gaia}/\textit{Gaia} & $25.00 \pm 0.26$ & $11.673 \pm 0.000$ & ZH & $12.572 \pm 0.002$ & VMC & $-0.008 \pm 0.047$ & $1.867 \pm 0.056$ & $0.699 \pm 0.060$ & LG1 \\
LLS~1\,716 & -- & BAT99~108 & 84.67665 & -69.10080 & WR* & WN5h & 12.205 & 0.345 & HST/HST & -- & -- & -- & -- & -- & -- & -- & -- & LG7 \\
LLS~1\,827 & -- & [DCL92]~205 & 84.91763 & -69.77194 & * & O5 & 12.071 & 0.982 & ZH/ZH & -- & -- & -- & -- & -- & -- & -- & -- & LG8 \\
LLS~5\,535 & 4658308373816333312 & LHA~120-S~101 & 80.50181 & -68.87201 & Em* & OBe & 13.191 & 0.356 & \textit{Gaia}/\textit{Gaia} & $160.07 \pm 0.01$ & $13.789 \pm 0.022$ & ZH & $12.039 \pm 0.002$ & VMC & $-0.001 \pm 0.023$ & $1.602 \pm 0.036$ & $0.216 \pm 0.036$ & LG1 \\
LLS~5\,875 & -- & Brey~58a & 83.92619 & -69.19794 & WR* & WC & 13.333 & 0.939 & 2MASS/2MASS & -- & -- & -- & $11.169 \pm 0.054$ & 2MASS & -- & -- & -- & LG7 \\
LLS~5\,936 & 4658617336550693376 & LHA~120-S~44 & 83.01431 & -67.70712 & Y*O & O9I[e] & 13.519 & 1.168 & \textit{Gaia}/\textit{Gaia} & $38.98 \pm 0.01$ & $13.369 \pm 0.059$ & ZH & $10.662 \pm 0.001$ & VMC & $-0.039 \pm 0.110$ & $1.677 \pm 0.138$ & $0.340 \pm 0.148$ & LG1 \\
LLS~12\,705 & 4657676360777000320 & VFTS~797 & 84.87768 & -69.15732 & * & O3.5V((n))((fc)) & 14.637 & 0.134 & \textit{Gaia}/\textit{Gaia} & $0.20 \pm 0.74$ & $13.823 \pm 0.038$ & ZH & $14.491 \pm 0.005$ & VMC & $0.015 \pm 0.031$ & $1.794 \pm 0.043$ & $0.503 \pm 0.045$ & LG2 \\
LLS~12\,990 & 4657637190669985408 & [FBM2009]~32 & 84.90477 & -69.72925 & * & B0Ia & 14.518 & 0.294 & \textit{Gaia}/\textit{Gaia} & $143.44 \pm 0.02$ & $13.667 \pm 0.066$ & ZH & $13.738 \pm 0.004$ & VMC & $-0.001 \pm 0.026$ & $1.868 \pm 0.037$ & $0.640 \pm 0.038$ & LG4 \\
LLS~13\,089 & -- & [HSH95]~89 & 84.67709 & -69.10086 & -- & -- & 14.685 & 0.451 & HST/HST & -- & -- & -- & -- & -- & -- & -- & -- & LG10 \\
LLS~22\,991 & 4657692028820011136 & 2MASS~J05373778-6905455 & 84.40751 & -69.09597 & SB* & O8.5V+B0.2:V & 15.067 & 0.109 & \textit{Gaia}/\textit{Gaia} & $0.02 \pm 0.12$ & $14.312 \pm 0.038$ & ZH & $14.755 \pm 0.007$ & VMC & $-0.006 \pm 0.030$ & $1.646 \pm 0.048$ & $0.624 \pm 0.044$ & LG3 \\
LLS~23\,285 & -- & -- & 84.16095 & -69.14861 & -- & -- & 15.642 & 0.140 & HST/HST & -- & $14.930 \pm 0.144$ & SM & -- & -- & -- & -- & -- & LG11 \\
LLS~24\,282 & 4657947626577977984 & SV*~HV~12326 & 80.86120 & -69.98036 & s*r & M6.5Ib & 15.583 & 3.674 & \textit{Gaia}/\textit{Gaia} & -- & $17.621 \pm 0.051$ & ZH & $10.575 \pm 0.001$ & VMC & $0.106 \pm 0.100$ & $2.346 \pm 0.142$ & $-0.000 \pm 0.132$ & LG7 \\
LLS~34\,793 & 4659472172498144640 & UCAC4~114-009339 & 84.95195 & -67.38986 & Em* & B2Ve & 16.292 & 0.024 & \textit{Gaia}/\textit{Gaia} & $174.19 \pm 0.11$ & $15.536 \pm 0.030$ & ZH & $16.071 \pm 0.014$ & VMC & $-0.028 \pm 0.046$ & $1.496 \pm 0.057$ & $0.526 \pm 0.065$ & LG4 \\
LLS~38\,090 & 4654870888108125696 & LHA~120-S~146 & 70.75815 & -70.56813 & Y*O & O9V & 16.163 & 0.377 & \textit{Gaia}/\textit{Gaia} & -- & $14.532 \pm 0.075$ & ZH & $13.950 \pm 0.098$ & 2MASS & $-0.015 \pm 0.087$ & $2.068 \pm 0.089$ & $-0.250 \pm 0.105$ & LG4 \\
LLS~38\,600 & 4658829061266063232 & BAT99~23 & 78.48348 & -67.41014 & WR* & WN3(h) & 16.965 & 0.763 & \textit{Gaia}/\textit{Gaia} & $15.03 \pm 0.93$ & $17.077 \pm 0.041$ & ZH & $14.937 \pm 0.007$ & VMC & $0.007 \pm 0.055$ & $1.732 \pm 0.069$ & $0.180 \pm 0.075$ & LG5 \\
LLS~40\,002 & 4657685569167247104 & -- & 84.68145 & -69.08340 & -- & -- & 17.026 & -0.346 & \textit{Gaia}/\textit{Gaia} & -- & $15.569 \pm 0.139$ & ZH & -- & -- & $0.095 \pm 0.101$ & $1.811 \pm 0.177$ & $0.805 \pm 0.136$ & LG11 \\
LLS~40\,422 & -- & Brey~78 & 84.66766 & -69.09977 & WR* & O2.5If*/WN6 & 17.861 & 0.186 & HST/HST & -- & -- & -- & -- & -- & -- & -- & -- & LG7 \\
LLS~40\,838 & 4657679521872856320 & OGLE~LMC-ECL-21437 & 84.67753 & -69.13298 & EB* & B1V & 17.164 & 0.583 & \textit{Gaia}/\textit{Gaia} & -- & $17.051 \pm 0.099$ & ZH & $17.882 \pm 0.115$ & VMC & $0.003 \pm 0.091$ & $1.733 \pm 0.133$ & $0.767 \pm 0.127$ & LG9 \\
\hline
SLS~127 & 4686487722908882048 & LHA~115-S~64 & 22.82664 & -72.82164 & Em* & OBe & 12.377 & 0.198 & \textit{Gaia}/\textit{Gaia} & $35.72 \pm 0.05$ & $12.764 \pm 0.014$ & SM & $11.759 \pm 0.001$ & VMC & $0.047 \pm 0.035$ & $1.148 \pm 0.045$ & $-1.234 \pm 0.044$ & SG1 \\
SLS~147 & 4687207593802365568 & RMC~43 & 17.23687 & -73.04289 & sg* & A3Ia & 11.311 & 0.260 & \textit{Gaia}/\textit{Gaia} & $9.96 \pm 0.80$ & $12.703 \pm 0.039$ & SM & $10.858 \pm 0.001$ & VMC & $-0.003 \pm 0.027$ & $0.853 \pm 0.040$ & $-1.193 \pm 0.035$ & SG1 \\
SLS~464 & 4685944323714595072 & BBB~SMC~206 & 12.11259 & -73.20342 & s*r & M0Ia & 11.417 & 2.174 & \textit{Gaia}/\textit{Gaia} & $159.74 \pm 0.04$ & $15.854 \pm 0.060$ & SM & $8.055 \pm 0.000$ & VMC & $0.034 \pm 0.026$ & $0.702 \pm 0.038$ & $-1.140 \pm 0.037$ & SG1 \\
SLS~1\,007 & 4685925872530280192 & AzV~44 & 12.19763 & -73.41648 & EB* & O6.5V:nnn & 13.618 & -0.364 & \textit{Gaia}/\textit{Gaia} & $115.54 \pm 0.11$ & $12.911 \pm 0.038$ & ZH & $14.230 \pm 0.004$ & VMC & $-0.052 \pm 0.044$ & $0.584 \pm 0.054$ & $-1.221 \pm 0.054$ & SG1 \\
SLS~1\,073 & 4685919206739479296 & SK~100 & 15.37484 & -72.94067 & s*b & O9.5Ia & 13.524 & -0.267 & \textit{Gaia}/\textit{Gaia} & $181.48 \pm 0.06$ & $12.243 \pm 0.010$ & ZH & $13.881 \pm 0.004$ & VMC & $0.024 \pm 0.036$ & $0.701 \pm 0.047$ & $-1.311 \pm 0.041$ & SG1 \\
SLS~1\,174 & 4685879040157034624 & LIN~160 & 12.68095 & -73.45162 & WR* & WN6 & 13.581 & -0.136 & \textit{Gaia}/\textit{Gaia} & $20.70 \pm 0.58$ & $12.363 \pm 0.010$ & ZH & $12.757 \pm 0.002$ & VMC & $-0.085 \pm 0.140$ & $0.773 \pm 0.166$ & $-1.182 \pm 0.168$ & SG1 \\
SLS~3\,591 & 4685931438808874496 & SMC~AB~2 & 12.12853 & -73.26274 & WR* & WN4.5/O3 & 14.339 & -0.213 & \textit{Gaia}/\textit{Gaia} & $16.70 \pm 0.88$ & $13.233 \pm 0.030$ & ZH & $14.253 \pm 0.004$ & VMC & $0.009 \pm 0.033$ & $0.668 \pm 0.047$ & $-1.257 \pm 0.045$ & SG1 \\
SLS~3\,634 & -- & [HRZ99]~2 & 17.30429 & -73.19389 & * & O6.5 & 14.866 & -0.196 & ZH/ZH & -- & -- & -- & -- & -- & -- & -- & -- & SG8 \\
SLS~4\,415 & -- & LHA~115-N~77A & 15.70467 & -71.88833 & Em* & O9 & 14.399 & 0.305 & SM/SM & -- & $14.700 \pm 0.179$ & SM & $14.030 \pm 0.124$ & 2MASS & -- & -- & -- & SG7 \\
SLS~6\,654 & 4689015637641295104 & Cl*~NGC~346~NMC~34 & 14.75402 & -72.18592 & * & O6.5V & 15.454 & -0.315 & \textit{Gaia}/\textit{Gaia} & $3.65 \pm 2.89$ & $14.109 \pm 0.030$ & ZH & $16.384 \pm 0.016$ & VMC & $-0.030 \pm 0.038$ & $0.600 \pm 0.051$ & $-1.268 \pm 0.045$ & SG1 \\
SLS~8\,727 & 4689013472978130944 & [MA93]~1030 & 14.43463 & -72.26970 & Be* & B2(Be-Fe) & 15.256 & -0.058 & \textit{Gaia}/\textit{Gaia} & $179.92 \pm 0.08$ & $14.548 \pm 0.029$ & ZH & $14.968 \pm 0.006$ & VMC & $-0.051 \pm 0.032$ & $0.564 \pm 0.044$ & $-1.190 \pm 0.043$ & SG4 \\
SLS~9\,338 & 4688988282947346688 & RX~J0055.2-7238 & 13.82685 & -72.64779 & HXB & O9.5V & 15.919 & 0.253 & \textit{Gaia}/\textit{Gaia} & $30.32 \pm 0.24$ & $15.452 \pm 0.037$ & ZH & $14.882 \pm 0.121$ & 2MASS & $-0.053 \pm 0.051$ & $0.721 \pm 0.070$ & $-1.086 \pm 0.059$ & SG5 \\
SLS~16\,846 & 4685947038135411840 & [MA93]~172 & 11.84725 & -73.20761 & HXB & B1-B1.5V & 16.102 & 0.013 & \textit{Gaia}/\textit{Gaia} & $7.31 \pm 2.76$ & $15.139 \pm 0.044$ & ZH & $15.392 \pm 0.008$ & VMC & $0.044 \pm 0.038$ & $0.533 \pm 0.054$ & $-1.267 \pm 0.050$ & SG2 \\
SLS~16\,926 & 4689001206548001152 & [MFH2007]~SMC5-3919 & 14.27708 & -72.35840 & Be* & B1V & 16.684 & 0.022 & \textit{Gaia}/\textit{Gaia} & $0.96 \pm 1.96$ & $15.801 \pm 0.031$ & ZH & $16.552 \pm 0.017$ & VMC & $0.026 \pm 0.050$ & $0.720 \pm 0.070$ & $-1.273 \pm 0.061$ & SG5 \\
SLS~17\,660 & 4689015740720446976 & Cl*~NGC~346~MPG~477 & 14.77580 & -72.17634 & * & O9.5V & 16.363 & 0.396 & \textit{Gaia}/HST & -- & -- & -- & -- & -- & -- & -- & -- & SG7 \\
SLS~19\,165 & -- & Cl*~NGC~330~KWBBE~1041 & 13.90909 & -72.46530 & Be* & B1IV & 18.008 & -0.049 & ZH/ZH & -- & $17.401 \pm 0.061$ & ZH & $16.336 \pm 0.014$ & VMC & -- & -- & -- & SG7 \\
SLS~19\,175 & 4685942846246118528 & OGLE~SMC100.7~9155 & 11.96971 & -73.27611 & EB* & B0 & 17.244 & -0.022 & \textit{Gaia}/\textit{Gaia} & -- & $16.321 \pm 0.047$ & ZH & $18.524 \pm 0.154$ & VMC & $0.113 \pm 0.066$ & $0.634 \pm 0.082$ & $-1.303 \pm 0.087$ & SG7 \\
SLS~19\,215 & 4685881303675651072 & IRAS~00509-7342 & 13.16196 & -73.44003 & Y*O & O9.5 & 17.862 & 0.583 & \textit{Gaia}/\textit{Gaia} & -- & $16.800 \pm 0.203$ & ZH & $14.699 \pm 0.112$ & 2MASS & $0.161 \pm 0.534$ & $1.226 \pm 0.647$ & $-1.704 \pm 0.629$ & SG7 \\
\hline
BLS~15 & 4643498982924371968 & SkKM~372 & 34.25699 & -74.35110 & s?r & K/M & 13.451 & 1.563 & \textit{Gaia}/\textit{Gaia} & $5.37 \pm 0.17$ & $17.241 \pm 0.019$ & SM & $11.544 \pm 0.001$ & VMC & $0.039 \pm 0.020$ & $1.460 \pm 0.030$ & $-1.110 \pm 0.031$ & BG1 \\
BLS~27 & 4643665284057887104 & 2MASS~J02275501-7316205 & 36.97938 & -73.27240 & s?r & -- & 13.755 & 1.567 & \textit{Gaia}/\textit{Gaia} & $5.65 \pm 0.22$ & $17.376 \pm 0.036$ & SM & $10.882 \pm 0.001$ & VMC & $0.043 \pm 0.022$ & $1.605 \pm 0.034$ & $-1.032 \pm 0.032$ & BG1 \\
BLS~33 & 4643450673132036608 & [DI91]~1260 & 37.88389 & -73.72808 & * & -- & 13.927 & -0.286 & \textit{Gaia}/\textit{Gaia} & $8.36 \pm 1.91$ & $13.576 \pm 0.011$ & SM & $14.689 \pm 0.005$ & VMC & $-0.010 \pm 0.032$ & $1.602 \pm 0.041$ & $-0.986 \pm 0.044$ & BG1 \\
BLS~104 & 4642240866743866880 & AL~4 & 50.87057 & -72.72699 & Em* & -- & 14.664 & 0.010 & \textit{Gaia}/\textit{Gaia} & -- & $14.409 \pm 0.017$ & SM & $14.353 \pm 0.004$ & VMC & $-0.030 \pm 0.026$ & $2.023 \pm 0.038$ & $-0.598 \pm 0.037$ & BG5 \\
BLS~110 & 4643634360293415424 & IDK~A007 & 36.91284 & -73.76330 & * & B2V & 15.350 & -0.267 & \textit{Gaia}/\textit{Gaia} & $11.18 \pm 1.38$ & $15.164 \pm 0.014$ & SM & $15.904 \pm 0.010$ & VMC & $-0.116 \pm 0.190$ & $1.656 \pm 0.220$ & $-1.194 \pm 0.213$ & BG1 \\
BLS~116 & 4643344501540368640 & [DI91]~1127 & 36.62102 & -74.12932 & * & -- & 15.574 & -0.341 & \textit{Gaia}/\textit{Gaia} & $6.88 \pm 2.60$ & $15.218 \pm 0.014$ & SM & $16.308 \pm 0.013$ & VMC & $0.027 \pm 0.036$ & $1.622 \pm 0.055$ & $-1.107 \pm 0.043$ & BG1 \\
BLS~131 & 4643636898617662080 & IDK~A002 & 37.12385 & -73.67721 & * & B2 & 15.833 & -0.282 & \textit{Gaia}/\textit{Gaia} & $6.92 \pm 2.77$ & $15.643 \pm 0.014$ & SM & $16.124 \pm 0.012$ & VMC & $0.011 \pm 0.044$ & $1.585 \pm 0.052$ & $-1.026 \pm 0.054$ & BG1 \\
BLS~147 & -- & 2MASS~J03534174-7356023 & 58.42392 & -73.93399 & * & -- & 16.359 & -0.140 & SM/SM & -- & $16.381 \pm 0.026$ & SM & $16.424$ & 2MASS & -- & -- & -- & BG8 \\
BLS~162 & 4644497953661373440 & IDK~C001 & 35.11932 & -72.80253 & * & B2V & 16.638 & -0.186 & \textit{Gaia}/\textit{Gaia} & $1.07 \pm 2.08$ & $16.807 \pm 0.026$ & SM & $17.048 \pm 0.022$ & VMC & $0.002 \pm 0.049$ & $1.548 \pm 0.064$ & $-1.149 \pm 0.059$ & BG2 \\
\hline
\end{tabular}
\end{adjustbox}
\tablefoot{The Syn. source column lists the origin of $G_{\rm syn}$ and $(G_{\rm BP}-G_{\rm RP})_{\rm syn}$, respectively: \textit{Gaia} = \textit{Gaia}~DR3 \citep{Valletal23}; HST = \textit{HST} \citep{Crowetal16,Sabbetal16}; ZH = MCPS \citep{Zarietal02,Zarietal04}; SM = SkyMapper \citep{Onkeetal24}; and 2MASS = 2MASS \citep{Skruetal06}. $S_G$ is the \textit{Gaia} G-band variability statistic \citep{Maizetal23}. The adopted $U$ magnitude is from MCPS (ZH; \citealt{Zarietal02,Zarietal04}) when available and otherwise from SkyMapper (SM; \citealt{Onkeetal24}). The adopted $K_s$ magnitude is from VMC (VMC; \citealt{Cionetal11b}) when available and otherwise from 2MASS (2MASS; \citealt{Skruetal06}). Photometric and astrometric uncertainties are reported next to their measurements. Parallaxes and proper motions are the corrected \textit{Gaia} DR3 quantities. Otype and SpT are the SIMBAD object and spectral types. Missing values are denoted by a dash, and as $-99$ in the electronic catalogue.}
\end{sidewaystable*}

\end{appendix}


\end{document}